\documentclass[11pt]{article}

\usepackage{microtype}
\usepackage{mathtools}
\usepackage{float}
\usepackage{hhline}
\usepackage[math]{cellspace}
\usepackage{chet}

\newlength{\nodesep}
\newcommand{\CO}{\mathcal{O}}
\newcommand{\CT}{\mathcal{T}}

\newcommand{\CN}{\mathcal{N}}
\newcommand{\COb}{{\bar{\mathcal{O}}}}
\newcommand{\CTb}{{\bar{\mathcal{T}}}}
\newcommand{\COind}[1]{\CO_{#1_1\hspace{-0.8pt}\ldots#1_j;\,
  \dot{#1}_1\hspace{-0.8pt}\ldots\dot{#1}_\jb}}
\newcommand{\COindb}[1]{\COb_{#1_1\hspace{-0.8pt}\ldots#1_\jb;\,
  \dot{#1}_1\hspace{-0.8pt}\ldots\dot{#1}_j}}
\newcommand{\Qb}{{\bar{Q}}}
\newcommand{\qb}{{\bar{q}}}
\newcommand{\jb}{{\bar{\jmath}}\hspace{0.9pt}}
\newcommand{\thetab}{{\bar{\theta}}}

\newcommand{\alphad}{{\dot{\alpha}}}
\newcommand{\betad}{{\dot{\beta}}}
\newcommand{\xup}{{\text{x}}}
\newcommand{\vev}[1]{\langle #1\rangle}
\newcommand{\Vev}[1]{\left\langle #1\right\rangle}

\newcommand{\osbx}[2]{x_{#1\hspace{1pt}}{\!}^{#2}}

\date{July 2014}

\title{Two\hspace{1.5pt}-point functions of conformal primary operators\\
\vspace{5pt}
in $\mathcal{N}=1$ superconformal theories}

\author{Daliang Li and Andreas Stergiou
\emails{(\href{mailto:daliang.li@yale.edu}{daliang.li},
\href{mailto:andreas.stergiou@yale.edu}{andreas.stergiou})@yale.edu}}

\affiliation{Department of Physics, Yale University, New Haven, CT 06520
USA}

\abstract{In $\mathcal{N}=1$ superconformal theories in four dimensions the
form of two-point functions of superconformal multiplets is known up to an
overall constant. A superconformal multiplet contains several conformal
primary operators, whose two-point function coefficients can be determined
in terms of the multiplet's quantum numbers. In this paper we work out
these coefficients in full generality, i.e.\ for superconformal multiplets
that belong to any irreducible representation of the Lorentz group
with arbitrary scaling dimension and R-charge. From our results we recover
the known unitarity bounds, and also find all shortening conditions, even
in non-unitary theories.  For the purposes of our computations we have
developed a \emph{Mathematica} package for the efficient handling of
expansions in Grassmann variables.}

\begin{document}

\maketitle

\newsec{Introduction}
Conformal field theories (CFTs) in four dimensions are abundant and have
been studied extensively.  However, the full spectrum of consequences of
conformal symmetry has remained elusive. This has been made strikingly
clear by the recent developments in the numerical conformal bootstrap
approach, an implementation of the conformal
bootstrap~\cite{Polyakov:1974gs, Ferrara:1973yt} in higher dimensions
initiated in~\cite{Rattazzi:2008pe}.  The numerical bootstrap has uncovered
extremely interesting implications of conformal symmetry and unitarity. Its
success also signals the possible existence of more hidden constraints of
conformal symmetry yet to be discovered.

The main quantities of interest in CFTs, as in any quantum field theory,
are correlation functions.  More specifically, in CFTs the focus is on
correlation functions of primary operators, defined as the operators that
are annihilated by the generator of special conformal transformations at
the origin.  Conformal symmetry places powerful constraints on the form of
such correlation functions, see e.g.~\cite{Osborn:1993cr,
Erdmenger:1996yc}. For example, two-point functions are fixed up to an
overall constant, three-point functions are fixed up to a finite set of
constants, and four-point functions can be expressed as a sum in terms of
conformal blocks, whose explicit form has been worked out in some even
dimensions in~\cite{Dolan:2000ut, Dolan:2003hv}. The bootstrap program uses
these expressions, along with crossing symmetry and sophisticated numerical
analysis, to produce deep and model-independent results about the spectrum
of operators and their scaling dimensions in CFTs, as well as coefficients
in the operator product expansion.

The numerical bootstrap has also been applied to four-dimensional $\CN=1$
superconformal theories (SCFTs)~\cite{Poland:2010wg, Berkooz:2014yda}.
Progress, however, hinges on the understanding of superconformal
correlation functions; in particular, superconformal blocks appearing in
four-point functions. For cases involving chiral operators, linear
multiplets, or general scalar operators, the corresponding superconformal
blocks are known~\cite{Poland:2010wg, Fortin:2011nq, Fitzpatrick:2014oza,
Khandker:2014mpa}, which opens the door to the application of the conformal
bootstrap to these cases.

The form of correlation functions in SCFTs has been studied extensively
in~\cite{Park:1997bq, Osborn:1998qu, Park:1999pd} and also, using
superembedding methods, in \cite{Kuzenko:2006mv, Goldberger:2011yp,
Siegel:2012di, Kuzenko:2012tb, Goldberger:2012xb, Khandker:2012pa}, with
elegant results for the two- and three-point functions of superfields.
These results succinctly encapsulate the correlation functions of all the
different components of the superfield, including the conformal primaries.
However, for many physical applications it is necessary to work out
explicitly these component correlation functions and their relations as
imposed by superconformal symmetry. For example, to determine the
superconformal blocks as linear combinations of conformal blocks, it is
necessary to explicitly work out the relations between the two-point
function coefficients of the various conformal primary components of a
superconformal multiplet.

From the superconformal two- and three-point functions one can also extract
the operator product expansion, and use it to explore the phenomenology of
models of supersymmetry breaking with a superconformal hidden sector in the
ultraviolet~\cite{Fortin:2011ad, Fortin:2012tp, Kumar:2014uxa}.  In
addition to direct physical applications, the coefficients in the component
correlation functions expose rich structures of SCFTs. For example, in the
case of the two-point function, one can read off all possible shortening
conditions and unitarity bounds on the most general supermultiplet.

There is a systematic way to decompose the known form of a superfield
correlation function into component correlation functions. One expands each
participating superfield in terms of the Grassmann coordinates $\theta$,
$\bar{\theta}$, yielding, at each order, a linear combination of conformal
primary component operators and possible descendants. Since the correlation
functions among these are determined by conformal symmetry up to unknown
constants, one can reconstruct the superfield correlation function using
them, and then compare to the known form as follows from superconformal
invariance.  This comparison uniquely determines all unknown coefficients
in the component correlators in terms of the coefficients in the superfield
correlator and the quantum numbers of the superfield. In this work we apply
this method to the most general superconformal two-point function. We leave
the case of three-point functions for future work.

In practice, the computation outlined in the previous paragraph is rather
complicated. In order to compare the known and reconstructed forms of the
superconformal two-point function order by order in $\theta$,
$\bar{\theta}$, one first needs to expand its known form. This expansion is
already rather involved due to the various relations among Lorentz
covariant structures, such as Fierz identities. The contribution from the
various descendants to the reconstructed form poses another challenge, as
they involve complicated derivative operators acting on two-point functions
of conformal primaries.  In order to make the computation manageable, we
developed a \emph{Mathematica} package that can perform
$\theta$-expansions, simplify expressions with various rules satisfied by
four-vectors and spinors, and compute two-point functions involving
conformal descendants.

Although in this work we take a direct route in obtaining our results, it
is natural to ask if there is an alternative, less computationally
challenging way, to obtain the same answers. Although we don't have an
answer to this question, we believe that our results may help in the
development of a method of achieving the conformal decomposition of the
superconformal correlation functions with less computational effort. We
should note here that there exists another way to obtain the same results,
using radial quantization and the superconformal
algebra~\cite{Minwalla:1997ka}, but it is rather tedious as well.

This paper is organized as follows. In the next section we present the
construction of the irreducible Lorentz representations for the
superconformal descendants of a general superconformal primary operator
$\COind{\alpha}$. We also remind the reader of basic facts on $\CN=1$
superconformal representation theory. In section \ref{sumres} we summarize,
for the reader's convenience, our results for the various
superconformal-descendant but conformal-primary two-point functions. We
also make various comments on our results, and rederive the well-known
unitarity bounds~\cite{Flato:1983te, Dobrev:1985qv} and multiplet
shortening conditions. The latter are obtained in all generality without
imposing unitarity. In section \ref{TwoPFs} we provide further details on
the derivation of these results, including all the ingredients necessary
for constructing the conformal primary operators in a superconformal
multiplet. In section \ref{Example} we give equations for the supercurrent
supermultiplet, or Ferrara--Zumino multiplet~\cite{Ferrara:1974pz}, which
contains the R-current, the supersymmetry current, and the stress-energy
tensor. We summarize in section \ref{Summary} with comments on possible
uses of our results. In appendix \ref{EtaForm} we outline the method we
used for our computations, and in appendix \ref{Package} we provide more
details on our \emph{Mathematica} package.

We follow the conventions of Wess \& Bagger~\cite{Wess:1992cp}.

\newsec{\texorpdfstring{$\CN=1$}{N=1} superconformal primary operators and
their descendants}
Local operators in a CFT can be classified into representations of the
conformal algebra. One can regard the generator of translations, $P_{\mu}$,
as a raising operator, and the generator of special conformal
transformations, $K_{\mu}$, as a lowering operator. A representation can be
constructed by applying $P_{\mu}$ in all possible ways on a conformal
primary operator, which is annihilated by $K_{\mu}$ at the origin. The
operators obtained by acting with $P_{\mu}$ are called conformal
descendants. In four dimensions, a generic conformal primary operator can
be characterized by its scaling dimension $\Delta$ and its Lorentz
representation $(j/2,\jb/2)$, explicitly constructed as
$\COind{\alpha}(x)$. $\COind{\alpha}$ is assumed totally symmetric in its
dotted and, separately, its undotted indices,
\eqn{\COind{\alpha}=\CO_{(\alpha_{1}\hspace{-0.8pt}\ldots\alpha_j);\,
(\alphad_{1}\hspace{-0.8pt}\ldots\alphad_\jb\hspace{-0.7pt})}
\equiv\frac{1}{j!\jb!}\sum_{(I,J)=(1,1)}^{(j!,
\hspace{1pt}\jb!)}\mathcal{P}_{\alpha_{1}\hspace{-0.8pt}
\ldots\alpha_{j}}^{(I)}\mathcal{P}_{\dot{\alpha}_{1}
\hspace{-0.8pt}\ldots\dot{\alpha}_{\jb}}^{(J)}
\CO_{\alpha_{1}\hspace{-0.8pt}\ldots\alpha_{j};\,\dot{\alpha}_{1}
\hspace{-0.8pt}\ldots\dot{\alpha}_\jb},}[symm]
where $\mathcal{P}_{\alpha_{1}\hspace{-0.8pt}\ldots\alpha_{j}}^{(I)}$ runs
over all possible permutations of the undotted indices and similarly for
$\mathcal{P}_{\alphad_{1}\hspace{-0.8pt}\ldots\alphad_{\jb}}^{(J)}$ and the
dotted indices.  Constructed this way, $\CO$ furnishes an irreducible
representation (irrep) of the Lorentz group.  Obviously, $j$ and $\jb$ are
nonnegative integers.

The $\mathcal{N}=1$ superconformal algebra in four dimensions extends the
conformal algebra with the supercharges $Q_{\alpha}$, $\Qb_\alphad$, the
superconformal supercharges $S_{\alpha}$, $\bar{S}_{\dot{\alpha}}$, and the
$U(1)_{R}$ generator, $R$. One can regard $P_{\mu}$, $Q_{\alpha}$, and
$\bar{Q}_{\dot{\alpha}}$ as raising operators, and $K_{\mu}$, $S_{\alpha}$,
$\bar{S}_{\dot{\alpha}}$ as lowering operators. A generic representation
can be constructed by acting with raising operators on a superconformal
primary operator $\COind{\alpha}$, characterized by
$(j,\jb\hspace{-0.7pt},q,\bar{q})$, where $(j/2,\jb/2)$ again labels its
Lorentz representation, and the conformal weights $q$, $\bar{q}$ are
related to scaling dimension and $R$-charge by
\eqn{\Delta=q+\qb,\qquad R=\tfrac23(q-\qb).}[]
$\COind{\alpha}$ is bosonic if $j+\jb$ is even, and fermionic if $j+\jb$ is
odd. If $j=\jb$, then $\CO$ furnishes an irreducible (symmetric traceless)
integer-spin representation of the Lorentz group with spin $\ell=j=\jb$.
An operator with zero R-charge has $q=\qb=\Delta/2$. We will also consider
the conjugate of $\CO$, $\COb$, obtained via
$(j,\jb\hspace{-0.7pt},q,\qb)\to(\jb\hspace{-0.7pt},j,\qb,q)$.

\subsec{Components of \texorpdfstring{$\CO$}{O}}
One can construct the full superconformal multiplet by applying raising
operators $P_{\mu}$, $Q_{\alpha}$, and $\bar{Q}_{\dot{\alpha}}$ on
$\mathcal{O}$. Since $Q$ and $\bar{Q}$ are nilpotent, such a multiplet only
contains a finite number of conformal multiplets. For example, if we apply
$Q^{2}$, the result $Q^{2}\CO$ is a superconformal descendant.
Nevertheless, it is still a conformal primary with quantum numbers
$(j,\jb\hspace{-0.7pt},q-1,\bar{q}+2)$ (or $\Delta=q+\bar{q}+1$ and
$R=\tfrac{2}{3}(q-\bar{q})-2$).

If we apply a single $Q_{\alpha}$, then the result would fall into two
different irreducible representations, since
$(\frac{1}{2},0)\otimes(\frac{j}{2},\frac{\jb}{2})=(\frac{j-1}{2},
\frac{\jb}{2})\oplus(\frac{j+1}{2},\frac{\jb}{2})$,
where $(\frac{j+1}{2},\frac{\jb}{2})$ corresponds to symmetrizing the
additional index, i.e.\ $Q_{(\alpha}\CO_{\alpha_{1}\hspace{-0.8pt}
\ldots\alpha_{j});\,\dot{\alpha}_{1}\hspace{-0.8pt}\ldots
\dot{\alpha}_{\jb}}$, and $(\frac{j-1}{2},\frac{\jb}{2})$ corresponds to
antisymmetrizing, i.e.\ $Q^{\alpha}\CO_{\alpha\alpha_{1}\hspace{-0.8pt}
\ldots\alpha_{j-1};\,\dot{\alpha}_{1}\hspace{-0.8pt}\ldots
\dot{\alpha}_{\jb}}$. More explicitly, with the conventions of \symm one
can derive the identity
\eqn{Q_{\alpha}\COind{\alpha}=Q_{(\alpha}\CO_{\alpha_{1}\hspace{-0.8pt}
\ldots\alpha_{j});\,\dot{\alpha}_{1}\hspace{-0.8pt}\ldots
\dot{\alpha}_{\jb}}+\frac{j}{j+1}\epsilon_{\alpha(\alpha_{1}}Q^{\beta}
\CO_{|\beta\alpha_{2}\ldots\alpha_{j});\,\dot{\alpha}_{1}\hspace{-0.8pt}
\ldots\dot{\alpha}_{\jb}}.}[Qdecomp]
Since\foot{For the superconformal algebra we use the conventions
of~\cite[Appendix A]{Fortin:2011nq}.} $[K^\mu,Q_\alpha]=
-\sigma^\mu_{\alpha\alphad}\bar{S}^{\alphad}$, the two operators on the
right-hand side are conformal primaries, characterized by quantum numbers
$(j\pm1,\jb\hspace{-0.7pt},q-\frac{1}{2}, \bar{q}+1)$.  They can thus be
denoted unambiguously by $(Q\CO)_{j\pm1,\hspace{1pt}\jb}$.  Note that in
the second term in the right-hand side of \Qdecomp the index $\beta$ of
$\CO$ is exchanged with each of $\alpha_2,\ldots,\alpha_j$ in the
symmetrization, but not with $\alpha_1$. Explicitly,
\eqn{\epsilon_{\alpha(\alpha_1}Q^\beta\CO_{|\beta\alpha_2\ldots\alpha_j);
\,\alphad_1\hspace{-0.8pt}\ldots\alphad_\jb}=\frac{1}{j(j!)}\sum_{(I,i)=
(1,1)}^{(j!,\hspace{0.7pt}j)}\mathcal{P}^{(I)}_{\beta\alpha_1
\hspace{-0.8pt}\ldots\alpha_{i-1}\alpha_{i+1}\hspace{-0.8pt}
\ldots\alpha_{j}}\epsilon_{\alpha\alpha_{i}}Q^\beta
\CO_{\beta\alpha_{1}\hspace{-0.8pt}\ldots\alpha_{i-1}\alpha_{i+1}
\hspace{-0.8pt}\ldots\alpha_{j};\,\alphad_1\hspace{-0.8pt}\ldots
\alphad_{\jb}}.}[]

If we apply $Q\bar{Q}$ on $\CO$, then we get four different operators
characterized by $(j\pm1,\jb\pm1,q+\frac{1}{2},\bar{q}+\frac{1}{2})$:
\eqna{ Q_\alpha\Qb_\alphad\COind{\alpha}&=
 Q_{(\alpha}\Qb_{(\alphad}\CO_{\alpha_1
\hspace{-0.8pt}\ldots
\alpha_j);\,\alphad_1\hspace{-0.8pt}\ldots\alphad_\jb\hspace{-0.7pt})}\\
&\quad+\frac{j}{j+1}\epsilon_{\alpha(\alpha_1}Q^\beta
\Qb_{(\alphad}\CO_{|\beta\alpha_2\hspace{-0.5pt}\ldots\alpha_j);\,
\alphad_1\hspace{-0.8pt}\ldots\alphad_\jb\hspace{-0.7pt})}\\
&\quad+\frac{\jb}{\jb+1}\epsilon_{\alphad(\alphad_1}Q_{(\alpha}
\Qb^{\dot{\beta}}\CO_{\alpha_1\hspace{-0.8pt}\ldots\alpha_j);\,
|\dot{\beta}\alphad_2\hspace{-0.5pt}\ldots\alphad_\jb\hspace{-0.7pt})}\\
&\quad+\frac{j\jb}{(j+1)(\jb+1)}\epsilon_{\alpha(\alpha_1}
\epsilon_{\alphad(\alphad_1}
Q^\beta\Qb^{\dot{\beta}}\CO_{|\beta\alpha_2\hspace{-0.5pt}\ldots\alpha_j);
\,|\dot{\beta}\alphad_2\hspace{-0.5pt}\ldots\alphad_\jb
\hspace{-0.7pt})},}[QQbtheta]
where, of course, the dotted indices do not participate in the
symmetrization of the undotted ones and vice-versa. These four operators
can be denoted by $(Q\Qb\CO)_{j\pm1,\hspace{1pt}\jb\pm1}$. The second,
third, and fourth operator in the right-hand side of \QQbtheta only exist
if $j\ne0$, $\jb\ne0$, and $j\jb\ne0$ respectively. The operators in
\QQbtheta are not conformal primaries, since, on a superconformal primary,
$[K^\mu,Q_\alpha\Qb_\alphad]=\tfrac12
i(\sigma^\mu\bar{\sigma}^\nu\sigma^\rho)_{\alpha\alphad}M_{\nu\rho}
+(2iD-3R)\sigma^\mu_{\alpha\alphad}$, where $M_{\mu\nu}$ is the generator
of Lorentz transformations and $D$ that of dilatations.  Nevertheless,
since $[K_\mu,P_\nu]=2i(\eta_{\mu\nu}D-M_{\mu\nu})$, conformal primaries
can be extracted out of them by subtracting $P\CO$ with appropriate
coefficients, which we will work out explicitly. After this substraction we
will obtain four conformal primaries with different Lorentz
representations. We denote these primaries with a subscript ``$p$'', i.e.\
$(Q\Qb\CO)_{j\pm1,\hspace{1pt}\jb\pm1;\hspace{1pt}p}$.

At higher orders we find $Q^2\COind{\alpha}$ and $\Qb^2\COind{\alpha}$, or
$(Q^2\CO)_{j,\hspace{1pt}\jb}$ and $(\Qb^2\CO)_{j,\hspace{1pt}\jb}$, which
are already conformal primary Lorentz irreps. Furthermore, the action of
$Q^2\bar{Q}$ on $\COind{\alpha}$ produces two operators, namely
$(Q^2\Qb\CO)_{j,\hspace{1pt}\jb\pm1}$, as does the action of $\Qb^2Q$,
namely $(\Qb^2Q\CO)_{j\pm1,\hspace{1pt}\jb}$. Each of these operators
contains a conformal primary.  Finally, the operator
$Q^2\Qb^2\COind{\alpha}$, or $(Q^2\Qb^2\CO)_{j,\hspace{1pt}\jb}$, contains
a single conformal primary.

\begin{figure}[ht]
  \centering
  \includegraphics[scale=0.98]{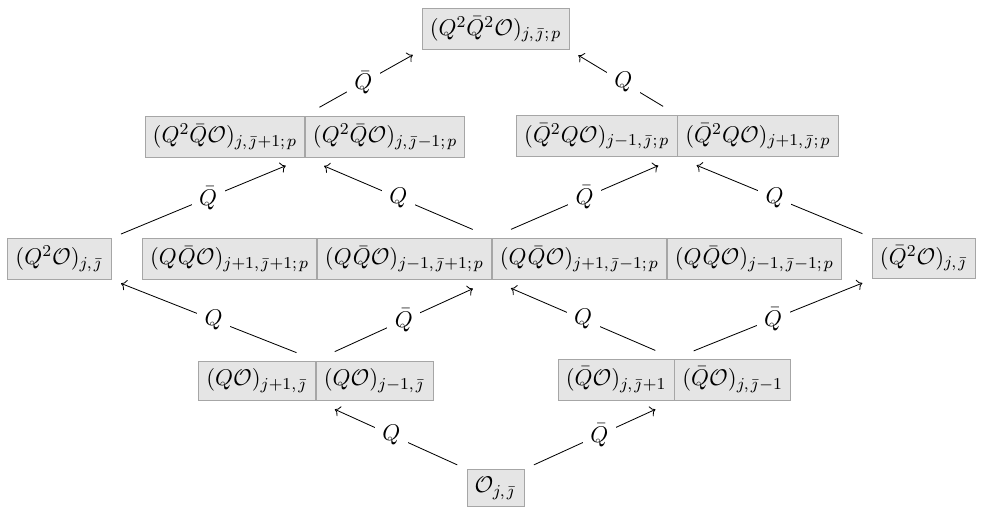}
  \caption{A superconformal multiplet consists of a finite number of
  conformal primary operators, related by supersymmetry.  Some of these
  primaries may become null when the multiplet's quantum numbers
  $(j,\jb,q,\qb)$ satisfy shortening conditions.} \label{fig:multiplet}
\end{figure}

We summarize the structure of an $\mathcal{N}=1$ superconformal
multiplet in Fig.~\ref{fig:multiplet}.

Generically, the multiplet contains sixteen different conformal primary
operators, but if the quantum numbers $(j,\jb\hspace{-0.7pt},q,\bar{q})$ obey
special conditions, then certain higher components may become null. We will
systematically determine all such conditions.

\newsec{Summary of results}[sumres]
For each of the sixteen conformal primary components mentioned above we
can write its two-point function with its conjugate as
\eqn{\vev{\CT_{\alpha_1\hspace{-0.8pt}\ldots\alpha_j;\,\alphad_1
\hspace{-0.8pt}\ldots\alphad_\jb}(x)
\CTb_{\smash{\beta_1\hspace{-0.8pt}\ldots\beta_\jb;
\,\betad_1\hspace{-0.8pt}\ldots\betad_j}}(0)}=C_{\CT}
\frac{\xup_{(\alpha_1\betad_1}\!\!\cdots
\xup_{\alpha_j)\betad_j}\xup_{\vphantom{\betad}
(\beta_1\alphad_1}\!\!\cdots \xup_{\vphantom{\betad}
\beta_\jb\hspace{-0.7pt})\alphad_\jb}}{x^{2(q_\CT^{\phantom{a}}+\qb_\CT^{\phantom{a}})
+j_\CT^{\phantom{a}}+\jb_\CT^{\phantom{a}}}},}[CTwoPF]
where $\xup_{\alpha\alphad}= \sigma^\mu_{\alpha\alphad} x_{\mu}$ and the
dotted indices do not participate in the symmetrization of the undotted
ones. In \CTwoPF, $\CT$ may be $\CO_{j,\hspace{1pt}\jb}$, $(Q\CO)_{j\pm1,
\hspace{1pt}\jb}$, etc.\ from Fig.~\ref{fig:multiplet}. In unitary
theories, $(-i)^{j_\CT^{\vphantom{a}}+\jb_\CT^{\vphantom{a}}}C_\CT>0$.

The $x$-dependence in \CTwoPF is uniquely determined by conformal symmetry,
while $C_{\CT}$ is determined by supersymmetry in terms of the coefficient
of the lowest component in Fig.~\ref{fig:multiplet}, $C_{\CO}$. The results
are summarized in Table~\ref{tab:Coeffs}.
\begin{table}[H]
  \centering
  \begin{tabular}[b]{>{$}Sr <{$}|*{1}{| >{$}Sl <{$}}}
    \CT & C_\CT \\
    \hhline{=::=}
    \CO_{j,\hspace{1pt}\jb}  & C_{\CO} \\
    \hhline{-||-}
    (Q\CO)_{j+1,\hspace{1pt}\jb} & 2iC_\CO\frac{j+2q}{(j+1)^2} \\
    \hhline{-||-}
    (Q\CO)_{j-1,\hspace{1pt}\jb} & 2iC_\CO\frac{(j+1)(j-2(q-1))}{j} \\
    \hhline{-||-}
    (\Qb\CO)_{j,\hspace{1pt}\jb+1} & 2iC_\CO\frac{\jb+2\qb}{(\jb+1)^2}
    \\
    \hhline{-||-}
    (\Qb\CO)_{j,\hspace{1pt}\jb-1} &
    2iC_\CO\frac{(\jb+1)(\jb-2(\qb-1))}{\jb} \\
    \hhline{-||-}
    (Q^2\CO)_{j,\hspace{1pt}\jb} & -2^4C_\CO(j+2q)(j-2(q-1)) \\
    \hhline{-||-}
    (\Qb^2\CO)_{j,\hspace{1pt}\jb} & -2^4C_\CO(\jb+2\qb)(\jb-2(\qb-1)) \\
    \hhline{-||-}
    (Q\Qb\CO)_{j+1,\hspace{1pt}\jb+1;\hspace{1pt}p} &
    -4C_\CO\frac{(j+2q) (\jb+2\qb) (j+\jb+2(q+\qb+1))}
    {(j+1)^2(\jb+1)^2(j+\jb+2(q+\qb))} \\
    \hhline{-||-}
    (Q\Qb\CO)_{j-1,\hspace{1pt}\jb+1;\hspace{1pt}p} &
    -4C_\CO\frac{(j+1)(j-2(q-1))(\jb+2\qb)(j-\jb-2(q+\qb))}
    {j(\jb+1)^2(j-\jb-2(q+\qb-1))}\\
    \hhline{-||-}
    (Q\Qb\CO)_{j+1,\hspace{1pt}\jb-1;\hspace{1pt}p} &
    -4C_\CO\frac{(\jb+1)(\jb-2(\qb-1))(j+2q)
    (j-\jb+2(q+\qb))}{\jb(j+1)^2(j-\jb+2(q+\qb-1))}\\
    \hhline{-||-}
    (Q\Qb\CO)_{j-1,\hspace{1pt}\jb-1;\hspace{1pt}p} &
    -4C_\CO\frac{(j+1)(\jb+1)(j-2(q-1))(\jb-2(\qb-1)) (j+\jb
    -2(q+\qb-1))}{j\jb(j+\jb-2(q+\qb-2))} \\
    \hhline{-||-}
    (Q^2\Qb\CO)_{j,\hspace{1pt}\jb+1;\hspace{1pt}p} &
    -2^5iC_\CO\frac{(j+2q)(j-2(q-1))(\jb+2\qb)
    (j+\jb+2(q+\qb+1))(j-\jb-2(q+\qb))}{(\jb+1)^2
    (j+\jb+2(q+\qb))(j-\jb-2(q+\qb-1))} \\
    \hhline{-||-}
    (Q^2\Qb\CO)_{j,\hspace{1pt}\jb-1;\hspace{1pt}p} &
    -2^5iC_\CO\frac{(\jb+1)(j+2q)(j-2(q-1))
    (\jb-2(\qb-1))(j+\jb-2(q+\qb-1))(j-\jb+2(q+\qb))}{\jb
    (j+\jb-2(q+\qb-2)) (j-\jb+2(q+\qb-1))} \\
    \hhline{-||-}
    (\Qb^2Q\CO)_{j+1,\hspace{1pt}\jb;\hspace{1pt}p} &
    -2^5iC_\CO\frac{(j+2q)(\jb+2\qb)(\jb-2(\qb-1))
    (j+\jb+2(q+\qb+1))(j-\jb+2(q+\qb))}{(j+1)^2
    (j+\jb+2(q+\qb))(j-\jb+2(q+\qb-1))}\\
    \hhline{-||-}
    (\Qb^2Q\CO)_{j-1,\hspace{1pt}\jb;\hspace{1pt}p} &
    -2^5iC_\CO\frac{(j+1)(\jb+2\qb)(j-2(q-1))
    (\jb-2(\qb-1))(j+\jb-2(q+\qb-1))(j-\jb-2(q+\qb))}{j
    (j+\jb-2(q+\qb-2)) (j-\jb-2(q+\qb-1))} \\
    \hhline{-||-}
    (Q^2\Qb^2\CO)_{j,\hspace{1pt}\jb;\hspace{1pt}p} &
    \begin{tabular}[l]{@{}c@{}}
      \hspace{-1.4cm}$2^8C_\CO\frac{(j+2q)(j-2(q-1))(\jb+2\qb)
      (\jb-2(\qb-1))(j-\jb+2(q+\qb))(j-\jb-2 (q+\qb))}
      {(j+\jb+2 (q+\qb))(j-\jb+2 (q+\qb-1))(j-\jb-2 (q+\qb-1))}$ \\
      \hspace{6cm}$\times\frac{(j+\jb+2 (q+\qb+1))(j+\jb-2 (q+\qb-1))}
      {(j+\jb-2 (q+\qb-2))}$
    \end{tabular}
  \end{tabular}
  \caption{The coefficients in \CTwoPF for the various operators $\CT$ from
  Fig.~\ref{fig:multiplet}. In our conventions $C_{\CO}=i^{j+\jb}c_{\CO}$,
  with $c_{\CO}>0$ in a unitary theory. $j$ and $\jb$ are non-negative
  integers. If $j=0$ (resp.\ $\jb=0$), the components labeled by $j-1$
  (resp.\ $\jb-1$) do not exist.} \label{tab:Coeffs}
\end{table}

\subsec{Comments}
As is well-known, in conformal theories two-point functions of primary
operators can be arranged in a diagonal basis with all coefficients set
equal to one by a proper choice of operator normalization.  In
superconformal theories, since supersymmetry relates the different
conformal primaries in the multiplet, their normalizations are fixed by
that of the lowest component.  For example, if the lowest component of a
scalar multiplet, $\CO$, is canonically-normalized, then the operator
$Q^2\CO$ is generally not. This is because the normalization of $Q$ is
fixed by the supersymmetry algebra,
$\{Q_\alpha,\Qb_\alphad\}=2\sigma^\mu_{\alpha\alphad}P_\mu$, due to the
fact that the normalization of $P_\mu$ is fixed by
$P_\mu\CO=i\partial_\mu\CO$. Of course, one can define the operator
$\tilde{\CO}=2^{-3}(q(q-1))^{-1/2}Q^2\CO$, which is canonically-normalized,
but then the required normalization coefficient is given exactly by our
results.  Furthermore, our results are important when one considers
three-point functions of conformal primary operators, where the operator
normalizations affect the overall coefficients of the three-point
functions.  This has been illustrated, for example, in the computations of
superconformal blocks we alluded to in the introduction.

From our results we can derive unitarity bounds~\cite{Flato:1983te,
Dobrev:1985qv} and all shortening conditions (noticing that $j,\jb\geq0$).
Requiring that all primary operators in the multiplet correspond to states
with positive norm results in the well-known unitarity bounds
\eqn{\begin{gathered}
  q=j=0,\quad \qb\geq\jb/2+1;\\
  \qb=\jb=0,\quad q\geq j/2+1;\\
  j,\jb\geq0,\quad q\geq j/2+1,\quad \qb\geq\jb/2+1,
\end{gathered}}[]
and also the trivial case $j=\jb=q=\qb=0$, corresponding to
the unit operator.  As the quantum numbers saturate the unitarity bounds,
the multiplet gets shortened as in Table \ref{tab:Shortening-Unitary}.
\begin{table}[H]
\centering
\begin{tabular}[b]{>{$}Sc <{$}|*{1}{| >{$}Sc <{$}}}
  \text{Short.\ Condition} & \text{Short Multiplet} \\
  \hhline{=::=}
  q=j=0 & \CO_{j,\hspace{1pt}\jb}\quad
  (\bar{Q}\CO)_{j,\hspace{1pt}\jb\pm1}\quad
  (\bar{Q}^{2}\CO)_{j,\hspace{1pt}\jb} \\
  \hhline{-||-}
  \qb=\jb=0 & \CO_{j,\hspace{1pt}\jb}\quad
  (Q\CO)_{j,\hspace{1pt}\jb\pm1}\quad
  (Q^{2}\CO)_{j,\hspace{1pt}\jb} \\
  \hhline{-||-}
  q=\frac{j}{2}+1 & \CO_{j,\hspace{1pt}\jb}\quad
  (Q\CO)_{j+1,\hspace{1pt}\jb}\quad
  (\bar{Q}\CO)_{j,\hspace{1pt}\jb\pm1}\quad
  (\Qb^2\CO)_{j,\hspace{1pt}\jb}\quad
  (Q\bar{Q}\CO)_{j+1,\hspace{1pt}\jb\pm1;\hspace{1pt}p}\quad
  (\Qb^2Q\CO)_{j+1,\hspace{1pt}\jb;\hspace{1pt}p}\\
  \hhline{-||-}
  \bar{q}=\frac{\jb}{2}+1 & \CO_{j,\hspace{1pt}\jb}\quad
  (Q\CO)_{j\pm1,\hspace{1pt}\jb}\quad (\bar{Q}\CO)_{j,\hspace{1pt}\jb+1}
  \quad
  (Q^2\CO)_{j,\hspace{1pt}\jb}\quad
  (Q\bar{Q}\CO)_{j\pm1,\hspace{1pt}\jb+1;\hspace{1pt}p}\quad
  (Q^2\Qb\CO)_{j,\hspace{1pt}\jb+1;\hspace{1pt}p}
  \end{tabular}\\
  \caption{Shortening conditions on a generic superconformal multiplet in
  unitary theories and the associated short multiplets. The intersection of
  short multiplets is taken if two corresponding shortening conditions are
  satisfied simultaneously.} \label{tab:Shortening-Unitary}
\end{table}

Actually, with our results we can obtain all shortening conditions on lowest weight superconformal multiplets in
non-unitary theories as well. We list these conditions and the
corresponding null components in Table \ref{tab:Shortening-Nonunitary}.
\begin{table}[ht]
\centering
\begin{tabular}[b]{>{$}Sc <{$}|*{1}{| >{$}Sc <{$}}}
  \text{Short.\ Condition} & \text{Null Components}\\
  \hhline{=::=}
  q=-\frac{j}{2} &
  \begin{tabular}[c]{@{}c@{}}
    $(Q\CO)_{j+1,\hspace{1pt}\jb}$\quad
    $(Q^{2}\CO)_{j,\hspace{1pt}\jb}$\quad
    $(Q\bar{Q}\CO)_{j+1,\hspace{1pt}\jb\pm1;\hspace{1pt}p}$\\
    $(Q^{2}\bar{Q}\CO)_{j,\hspace{1pt}\jb\pm1;\hspace{1pt}p}$\quad
    $(\Qb^{2}Q\CO)_{j+1,\hspace{1pt}\jb;\hspace{1pt}p}$\quad
    $(Q^{2}\bar{Q}^{2}\CO)_{j,\hspace{1pt}\jb;\hspace{1pt}p}$
  \end{tabular}\\
  \hhline{-||-}
    \bar{q}=-\frac{\jb}{2} &
  \begin{tabular}[c]{@{}c@{}}
    $(\bar{Q}\CO)_{j,\hspace{1pt}\jb+1}$\quad
    $(\bar{Q}^{2}\CO)_{j,\hspace{1pt}\jb}$\quad
    $(Q\bar{Q}\CO)_{j\pm1,\hspace{1pt}\jb+1;\hspace{1pt}p}$\\
    $(\Qb^2Q\CO)_{j\pm1,\hspace{1pt}\jb;\hspace{1pt}p}$\quad
    $(Q^2\Qb\CO)_{j,\hspace{1pt}\jb+1;\hspace{1pt}p}$\quad
    $(Q^{2}\bar{Q}^{2}\CO)_{j,\hspace{1pt}\jb;\hspace{1pt}p}$
  \end{tabular}\\
  \hhline{-||-}
  q=\frac{j}{2}+1 &
  \begin{tabular}[c]{@{}c@{}}
    $(Q\CO)_{j-1,\hspace{1pt}\jb}$\quad
    $(Q^{2}\CO)_{j,\hspace{1pt}\jb}$\quad
    $(Q\bar{Q}\CO)_{j-1,\hspace{1pt}\jb\pm1;\hspace{1pt}p}$\\
    $(Q^{2}\bar{Q}\CO)_{j,\hspace{1pt}\jb\pm1;\hspace{1pt}p}$\quad
    $(\Qb^{2}Q\CO)_{j-1,\hspace{1pt}\jb;\hspace{1pt}p}$\quad
    $(Q^{2}\bar{Q}^{2}\CO)_{j,\hspace{1pt}\jb;\hspace{1pt}p}$
  \end{tabular}\\
  \hhline{-||-}
  \qb=\frac{\jb}{2}+1 &
  \begin{tabular}[c]{@{}c@{}}
    $(\bar{Q}\CO)_{j,\hspace{1pt}\jb-1}$\quad
    $(\bar{Q}^{2}\CO)_{j,\hspace{1pt}\jb}$\quad
    $(Q\bar{Q}\CO)_{j\pm1,\hspace{1pt}\jb-1;\hspace{1pt}p}$\\
    $(\Qb^2Q\CO)_{j\pm1,\hspace{1pt}\jb;\hspace{1pt}p}$\quad
    $(Q^2\Qb\CO)_{j,\hspace{1pt}\jb-1;\hspace{1pt}p}$\quad
    $(Q^{2}\bar{Q}^{2}\CO)_{j,\hspace{1pt}\jb;\hspace{1pt}p}$
  \end{tabular}\\
  \hhline{-||-}
  q+\qb=\frac{j-\jb}{2} &
  (Q\Qb\CO)_{j-1;\hspace{1pt}\jb+1;\hspace{1pt}p}\quad
  (Q^2\Qb\CO)_{j,\hspace{1pt}\jb+1;\hspace{1pt}p}\quad
  (\Qb^2Q\CO)_{j-1,\hspace{1pt}\jb;\hspace{1pt}p}\quad
  (Q^2\Qb^2\CO)_{j,\hspace{1pt}\jb;\hspace{1pt}p}\\
  \hhline{-||-}
  q+\qb=\frac{\jb-j}{2} &
  (Q\Qb\CO)_{j+1;\hspace{1pt}\jb-1;\hspace{1pt}p}\quad
  (Q^2\Qb\CO)_{j,\hspace{1pt}\jb-1;\hspace{1pt}p}\quad
  (\Qb^2Q\CO)_{j+1,\hspace{1pt}\jb;\hspace{1pt}p}\quad
  (Q^2\Qb^2\CO)_{j,\hspace{1pt}\jb;\hspace{1pt}p}\\
  \hhline{-||-}
  q+\qb=\frac{j+\jb}{2}+1 &
  (Q\Qb\CO)_{j-1,\hspace{1pt}\jb-1;\hspace{1pt}p}\quad
  (Q^2\Qb\CO)_{j,\hspace{1pt}\jb-1;\hspace{1pt}p}\quad
  (\Qb^2Q\CO)_{j-1,\hspace{1pt}\jb;\hspace{1pt}p}\quad
  (Q^2\Qb^2\CO)_{j,\hspace{1pt}\jb;\hspace{1pt}p}\\
  \hhline{-||-}
  q+\qb=-\frac{j+\jb}{2}-1 &
  (Q\Qb\CO)_{j+1,\hspace{1pt}\jb+1;\hspace{1pt}p}\quad
  (Q^2\Qb\CO)_{j,\hspace{1pt}\jb+1;\hspace{1pt}p}\quad
  (\Qb^2Q\CO)_{j+1,\hspace{1pt}\jb;\hspace{1pt}p}\quad
  (Q^2\Qb^2\CO)_{j,\hspace{1pt}\jb;\hspace{1pt}p}
\end{tabular}
\caption{Shortening conditions and the associated null components of a
generic superconformal multiplet in a non-unitary SCFT.}
\label{tab:Shortening-Nonunitary}
\end{table}

Now, by inspection of the coefficients in Table~\ref{tab:Coeffs}, it
appears that some two-point functions diverge at certain values of
$(j,\jb\hspace{-0.7pt},q,\qb)$ consistent with the unitarity bounds.
Further inspection shows that this actually does not happen due to the
numerators also becoming zero, faster, in fact, than the denominators. To
make this more clear, let us consider a specific example.  For a scalar
operator $\CO$ it can be seen from our results\foot{The case of a scalar
$\CO$ was also worked out in \cite[Appendix A]{Kumar:2014uxa}.} that
\eqn{\langle(Q^2\bar{Q}^2 \CO)_p(x)(Q^2\Qb^2 \CO)_p^\dagger(0)\rangle =
2^{12}C_\CO\frac{q\bar{q}(q-1)(\bar{q}-1)(q+\bar{q})(q+\bar{q}+1)}
{(q+\bar{q}-1)(q+\bar{q}-2)}\frac{1}{x^{2(q+\bar{q}+2)}},}[QsqQbsqTP]
where
\eqn{(Q^2\bar{Q}^2\CO)_p=Q^2\bar{Q}^2\CO-2^4\frac{\bar{q}(\bar{q}-1)}
{(q+\bar{q}-1)(q+\bar{q}-2)}P^2\CO -
2^3\frac{\bar{q}-1}{q+\bar{q}-2}Q\text{P}\bar{Q}\CO.}[QsqQbsqPrim]
The two-point function \QsqQbsqTP diverges at $q+\qb=1$, unless $q=\qb-1=0$
or $q-1=\qb=0$. Additionally, unless $q=\qb=1$, \QsqQbsqTP diverges at
$q+\qb=2$. The well-defined cases just mentioned correspond to antichiral,
chiral, and linear multiplets respectively. They are the only cases
consistent with unitarity for which $q+\qb=1,2$, and so the two-point
function \QsqQbsqTP and the primary operator \QsqQbsqPrim are always
well-defined in a unitary theory. Note that, in non-unitary theories,
two-point functions may actually diverge, but then the primary operator
appearing in them is not well-defined. This can be easily seen from the
example above, since both \QsqQbsqTP and \QsqQbsqPrim diverge at the same
$q$, $\qb$. Hence, the divergences found in some primary two-point
functions in non-unitary theories are a reflection of the fact that the
associated primary operators cannot be defined.

\newsec{Two-point functions}[TwoPFs]
In this section we explicitly demonstrate the matching procedure we used to compute the two-point function coefficients of component primary operators.

It is convenient to define the supersymmetric interval between points $x_i$
and $x_j$,
\eqn{x_{\bar{\imath}j}=-x_{j\bar{\imath}}\equiv x_{ij}
-i\theta_i\sigma\bar{\theta}_i
-i\theta_j\sigma\bar{\theta}_j+2i\theta_j\sigma\bar{\theta}_i,
}[superdistance]
where $x_{ij}=x_i-x_j$. The notation $x_{\bar{\imath}j}$ indicates that
this quantity is antichiral at $z_i$ and chiral at
$z_j$~\cite{Osborn:1998qu}, where $z=(x,\theta,\thetab)$ is a point in
superspace.  The two-point function of a superconformal primary operator
with its conjugate can be written down very succinctly with the help of
\superdistance:
\eqn{\vev{\CO_{\alpha_1\hspace{-0.8pt}\ldots\alpha_j;\,\alphad_1
\hspace{-0.8pt}\ldots\alphad_\jb}(z_1)
\COb_{\smash{\beta_1\hspace{-0.8pt}\ldots\beta_\jb;\,\betad_1
\hspace{-0.8pt}\ldots\betad_j}}(z_2)}=C_\CO
\frac{\xup_{1\bar{2}(\alpha_1\betad_1}\!\!\cdots
\xup_{1\bar{2}\,\alpha_j)\betad_j}\xup_{\bar{1}2\vphantom{\betad}
(\beta_1\alphad_1}\!\!\cdots \xup_{\bar{1}2\,\vphantom{\betad}
\beta_\jb\hspace{-0.7pt})\alphad_\jb}}{\osbx{1\bar{2}}{2q+j}
\osbx{\bar{1}2}{2\qb+\jb}}.}[SCTwoPF]
For the coefficient $C_\CO$ in \SCTwoPF we may write
\eqn{C_{\CO}=i^{j+\jb} c_\CO, \qquad c_\CO>0\text{ in a unitary theory.}
}[TwoPFCoeff]
In a unitary theory we can of course always choose a basis for the nonzero
operators $\CO$ such that $c_{\CO}=1$, but we do not make this choice here.

The superfield $\CO(z)$ can be obtained by applying $e^{i\theta
Q+i\bar{\theta}\bar{Q}}$ on its zero component $\CO(x)$,
\eqn{\COind{\alpha}(z)=e^{i\theta
Q+i\thetab\Qb}\COind{\alpha}(x).}[]
Note that we use the symbol $\CO$ both for the superfield operator and its
zero component.  The Baker--Campbell--Hausdorff formula and the
supersymmetry algebra imply that
\eqn{e^{i\theta Q+i\bar{\theta}\bar{Q}}=e^{i\theta
Q}e^{i\bar{\theta}\bar{Q}}e^{\theta\text{P}\bar{\theta}},}[]
and expanding the exponentials it is straightforward to evaluate
\eqna{e^{i\theta Q+i\bar{\theta}\bar{Q}}&=1+i\theta Q+i\bar{\theta}\bar{Q}
+\tfrac12\theta\sigma^\mu\bar{\theta}(Q\sigma_\mu\bar{Q}+2P_\mu)
+\tfrac14\theta^2Q^2+\tfrac14\bar{\theta}^{\hspace{0.5pt}2}\bar{Q}^2\\
&\quad-\tfrac{i}{4}\theta^2\bar{\theta}^{\dot{\alpha}}(Q^2
\bar{Q}_{\dot{\alpha}}-2Q^\alpha\sigma^\mu_{\alpha\dot{\alpha}}P_\mu)
+\tfrac{i}{4}\bar{\theta}^{\hspace{0.5pt}2}\theta^\alpha(\bar{Q}^2Q_\alpha
+2\sigma^\mu_{\alpha\dot{\alpha}}\bar{Q}^{\dot{\alpha}}P_\mu)\\
&\quad+\tfrac{1}{2^4}\theta^2\bar{\theta}^{\hspace{0.5pt}2}(Q^2\bar{Q}^2
-4P^2-4Q\sigma^\mu\bar{Q}P_\mu).}[exponExp]

Our task is now straightforward: we need to perform the $\theta$-expansion
of both sides of \SCTwoPF, and read off the various two-point functions of
the conformal primary operators that appear in the expansion of the
left-hand side, which we can obtain from \exponExp. In practice, even
obtaining the $\theta$-expansion of the right-hand side of \SCTwoPF is a
very cumbersome computation, as can be seen from \superdistance, but,
fortunately, it can be coded, for example in \emph{Mathematica}. This
amounts to implementing spinors, 4-vectors and various relations between them, such as Fierz identities.  For more details on this
and other aspects of the computation the reader is referred to
appendix \ref{EtaForm}.

The other hurdle in this computation is the contamination of descendants
starting at order $\theta_1\thetab_1\theta_2\thetab_2$. Indeed, the
expansion of the right-hand side of \SCTwoPF contains contributions of
two-point functions involving descendants---for example a two-point
function of the form $\vev{P\CO(x)P^\dagger\COb(0)}$ at order
$\theta_1\thetab_1 \theta_2\thetab_2$. Such contamination has to be
appropriately subtracted out, by working out the linear combinations of
operators that are conformal primary. This can be done using information
from contributions to the right-hand side of \SCTwoPF coming purely from
two-point functions involving descendants. An example is the order
$\theta_1\thetab_1$, which is present in \SCTwoPF simply because of
two-point functions of the form $\vev{P\CO(x)\COb(0)}$.

In the remainder of this section we list our results for the various
primary two-point functions. Our method of computation is explained in
appendix \ref{EtaForm}.

\subsec{Orders \texorpdfstring{$\theta_1\thetab_2$}{theta1 thetabar2} and
\texorpdfstring{$\thetab_1\theta_2$}{thetabar1 theta2}}
For the symmetric part of $Q\CO$ we get
\eqna{\vev{Q_{(\alpha}\CO_{\alpha_1\hspace{-0.8pt}\ldots\alpha_j);\,
\alphad_1\hspace{-0.8pt}\ldots\alphad_\jb}(x) \Qb_{\smash{(\betad}}
\COb_{\smash{\beta_1\hspace{-0.8pt}\ldots\beta_\jb;\,\betad_1
\hspace{-0.8pt}\ldots\betad_j)}}(0)}&=\\
&\hspace{-2cm}2iC_\CO(-1)^{j+\jb+1}\frac{j+2q}{(j+1)^2}
\frac{\xup_{(\alpha\betad}\xup_{\alpha_1\betad_1}\!\!\cdots
\xup_{\alpha_j)\betad_j}\xup_{\vphantom{\betad}(\beta_1\alphad_1}\!\!\cdots
\xup_{\vphantom{\betad}\beta_\jb\hspace{-0.7pt})\alphad_{\jb}}}
{x^{2(q+\qb+1)+j+\jb}},}[symQO]
while for the antisymmetric part of $Q\CO$ we find
\eqna{\vev{Q^\alpha\CO_{\alpha\alpha_1\hspace{-0.8pt}\ldots\alpha_{j-1};\,
\alphad_1\hspace{-0.8pt}\ldots\alphad_\jb}(x)\Qb^\betad\COb_{\smash{\beta_1
\hspace{-0.8pt}\ldots\beta_\jb;\,\betad\betad_1
\hspace{-0.8pt}\ldots\betad_{j-1}}}(0)}&=\\
&\hspace{-4.2cm}2iC_\CO(-1)^{j+\jb+1}\frac{(j+1)(j-2(q-1))}{j}
\frac{\xup_{(\alpha_1\betad_1}
\!\!\cdots \xup_{\alpha_{j-1})\betad_{j-1}}\xup_{\vphantom{\betad}(\beta_1
\alphad_1}\!\!\cdots\xup_{\vphantom{\betad}\beta_\jb
\hspace{-0.7pt})\alphad_\jb}}{x^{2(q+\qb)+j+\jb}}.}[antisymQO]
As a consistency check on \antisymQO we see that for $j=1$ and $\jb=0$ the
corresponding two-point function, namely $\vev{Q^\alpha\CO_\alpha(x)
\Qb^\betad\COb_\betad(0)}=\vev{Q^\alpha\CO_\alpha(x)
(Q^\beta\CO_\beta)^\dagger(0)}$, is indeed positive in a unitary theory
(recall \TwoPFCoeff).

We also find
\eqna{\vev{\Qb_{(\alphad}\CO_{\alpha_1\hspace{-0.8pt}\ldots\alpha_j;
\,\alphad_1\hspace{-0.8pt}\ldots
\alphad_\jb\hspace{-0.7pt})}(x) Q_{\smash{(\beta}}
\COb_{\smash{\beta_1\hspace{-0.8pt}\ldots\beta_\jb\hspace{-0.7pt});\,
\betad_1\hspace{-0.8pt}\ldots\betad_j}}(0)}&=\\
&\hspace{-2cm}2iC_\CO(-1)^{j+\jb+1}\frac{\jb+2\qb}{(\jb+1)^2}
\frac{\xup_{(\alpha_1\betad_1}\!\!\cdots\xup_{\alpha_j)\betad_j}
\xup_{(\beta\alphad\vphantom{\betad}}\xup_{\vphantom{\betad}
\beta_1\alphad_1}\!\!\cdots\xup_{\vphantom{\betad}\beta_\jb
\hspace{-0.7pt})\alphad_{\jb}}}{x^{2(q+\qb+1)+j+\jb}},}[symQOp]
and
\eqna{\vev{\Qb^\alphad\CO_{\alpha_1\hspace{-0.8pt}\ldots\alpha_j;\,
\alphad\alphad_1\hspace{-0.8pt}\ldots\alphad_{\jb-1}}(x)Q^\beta
\COb_{\smash{\beta\beta_1\hspace{-0.8pt}\ldots\beta_{\jb-1};
\,\betad_1\hspace{-0.8pt}\ldots\betad_j}}(0)}&=\\
&\hspace{-4cm}2iC_\CO(-1)^{j+\jb+1}\frac{(\jb+1)(\jb-2(\qb-1))}{\jb}
\frac{\xup_{(\alpha_1\betad_1}\!\!\cdots \xup_{\alpha_j)\betad_j}
\xup_{\vphantom{\betad}(\beta_1\alphad_1}\!\!\cdots\xup_{\vphantom{\betad}
\beta_{\jb-1})\alphad_{\jb-1}}}{x^{2(q+\qb)+j+\jb}}.}[antisymQOp]
These, of course, can also be obtained from \symQO and \antisymQO with
$(j,\jb\hspace{-0.7pt},q,\qb)\to(\jb\hspace{-0.7pt},j,\qb,q)$.

\subsec{Orders \texorpdfstring{$\theta_1^2\thetab_2^{\hspace{0.7pt}2}$}
{theta1sq thetabar2sq} and \texorpdfstring{$\thetab_1^{\hspace{0.7pt}2}
\theta_2^{2}$}{thetabar1sq theta2sq}}
At orders $\theta_1^2\thetab_2^{\hspace{0.7pt}2}$ and
$\thetab_1^{\hspace{0.7pt}2}\theta_2^2$ we find
\eqna{\vev{Q^2\CO_{\alpha_1\hspace{-0.8pt}\ldots\alpha_j;\,\alphad_1
\hspace{-0.8pt}\ldots\alphad_\jb}(x)
\Qb^2\COb_{\smash{\beta_1\hspace{-0.8pt}\ldots\beta_\jb;\,
\betad_1\hspace{-0.8pt}\ldots\betad_j}}(0)}&=\\
&\hspace{-1.5cm}2^4C_\CO(j+2q)(j-2(q-1))\frac{
\xup_{(\alpha_1\betad_1}\!\!\cdots\xup_{\alpha_j)\betad_j}
\xup_{\vphantom{\betad}(\beta_1\alphad_1}\!\!\cdots
\xup_{\vphantom{\betad}\beta_\jb\hspace{-0.7pt})\alphad_{\jb}}}
{x^{2(q+\qb+1)+j+\jb}},}[QsqO]
and
\eqna{\vev{\Qb^2\CO_{\alpha_1\hspace{-0.8pt}\ldots\alpha_j;\,\alphad_1
\hspace{-0.8pt}\ldots\alphad_\jb}
(x)Q^2\COb_{\smash{\beta_1\hspace{-0.8pt}\ldots\beta_\jb;\,
\betad_1\hspace{-0.8pt}\ldots\betad_j}}(0)}&=
\\ &\hspace{-1.5cm}2^4C_\CO(\jb+2\qb)(\jb-2(\qb-1))\frac{
\xup_{(\alpha_1\betad_1}\!\!\cdots\xup_{\alpha_j)\betad_j}
\xup_{\vphantom{\betad}(\beta_1\alphad_1}\!\!\cdots
\xup_{\vphantom{\betad}\beta_\jb\hspace{-0.7pt})\alphad_{\jb}}}
{x^{2(q+\qb+1)+j+\jb}},}[QsqOp]
respectively. The overall sign here can be checked for a scalar operator,
taking into account the relation $\Qb^2\COb=-(Q^2\CO)^\dagger$, which
follows from the fact that the bosonic operator $Q^2$ acts with the adjoint
action, i.e.\ $Q^2\CO\equiv[Q^2,\CO]$.

\subsec{Order \texorpdfstring{$\theta_1\thetab_1\theta_2\thetab_2$}{theta1
thetabar1 theta2 thetabar2}}
At order $\theta_1\thetab_1\theta_2\thetab_2$ we have to consider
conformal descendant contributions to \SCTwoPF. More specifically, we can
write
\eqna{e^{i\theta Q+i\thetab \Qb}&\COind{\alpha}\big|_{\theta\thetab}=
-\theta^\alpha\thetab^\alphad\Big\{\Big[Q_{(\alpha}
\Qb_{(\alphad}\CO_{\alpha_1\hspace{-0.8pt}\ldots\alpha_j);
\,\alphad_1\hspace{-0.8pt}\ldots\alphad_\jb\hspace{-0.7pt})}\Big]_p-
c_1\text{P}_{\!(\alpha(\alphad}\CO_{\alpha_1\hspace{-0.8pt}\ldots\alpha_j);
\,\alphad_1\hspace{-0.8pt}\ldots\alphad_\jb\hspace{-0.7pt})}\Big\}\\
&\quad+\frac{j}{j+1}\theta_{(\alpha_1}\thetab^\alphad
\Big\{\Big[Q^\alpha\Qb_{(\alphad}
\CO_{|\alpha\alpha_2\hspace{-0.5pt}\ldots\alpha_j);\,\alphad_1
\hspace{-0.8pt}\ldots\alphad_\jb\hspace{-0.7pt})}\Big]_p
-c_2\text{P}^\alpha{}_{\!(\alphad}\CO_{|\alpha\alpha_2\hspace{-0.5pt}
\ldots\alpha_j);\,\alphad_1\hspace{-0.8pt}\ldots\alphad_\jb
\hspace{-0.7pt})}\Big\}\\
&\quad+\frac{\jb}{\jb+1}\theta^\alpha\thetab_{(\alphad_1}\Big\{
\Big[Q_{(\alpha}\Qb^{\alphad}\CO_{\alpha_1\hspace{-0.8pt}\ldots\alpha_j);\,
|\alphad\alphad_2\hspace{-0.5pt}\ldots\alphad_\jb}\Big]_p
+c_3\text{P}_{\!(\alpha}{}^{\!\alphad}\CO_{\alpha_1\hspace{-0.8pt}
\ldots\alpha_j);\,|\alphad\alphad_2\hspace{-0.5pt}\ldots\alphad_\jb
\hspace{-0.7pt})}\Big\}\\
&\quad-\frac{j\jb}{(j+1)(\jb+1)}\theta_{(\alpha_1}\thetab_{(\alphad_1}
\Big\{\Big[Q^\alpha\Qb^\alphad\CO_{|\alpha\alpha_2\hspace{-0.5pt}
\ldots\alpha_j);\,|\alphad\alphad_2\hspace{-0.5pt}\ldots\alphad_\jb
\hspace{-0.7pt})}\Big]_p-c_4\text{P}^{\alphad\alpha}
\CO_{|\alpha\alpha_2\hspace{-0.5pt}\ldots\alpha_j);\,|\alphad\alphad_2
\hspace{-0.5pt}\ldots\alphad_\jb\hspace{-0.7pt})}\Big\},}[]
where $[\,\cdot\,]_p$ denotes a conformal primary operator.  The
$\theta_1\thetab_1$ order of \SCTwoPF arises purely because of the
descendants above via two-point functions of the form
$\vev{P\CO(x)\COb(0)}$, and this allows us to compute
\begin{equation}
  \begin{aligned}
    c_1&=\frac{j-\jb+2(q-\qb)}{(j+1)(\jb+1)(j+\jb+2(q+\qb))},&\qquad
    c_2&=\frac{j+\jb-2(q-\qb-1)}{j(\jb+1)(j-\jb-2(q+\qb-1))},\\
    c_3&=-\frac{j+\jb+2(q-\qb+1)}{\jb(j+1)(j-\jb+2(q+\qb-1))}, &\qquad
    c_4&=\frac{j-\jb-2(q-\qb)}{j\jb(j+\jb-2(q+\qb-2))}.
\end{aligned}
  \label{QQbcs}
\end{equation}

With these results and \exponExp it is now easy to find the combinations of
the operators $Q\Qb\CO$ and $P\CO$ that are conformal primaries. For
example,
\eqn{\Big[Q_{(\alpha}\Qb_{(\alphad}\CO_{\alpha_1\hspace{-0.8pt}
\ldots\alpha_j);\,\alphad_1\hspace{-0.8pt}\ldots\alphad_\jb
\hspace{-0.7pt})}\Big]_p=Q_{(\alpha}\Qb_{(\alphad}\CO_{\alpha_1
\hspace{-0.8pt}\ldots\alpha_j);\,\alphad_1\hspace{-0.8pt}\ldots
\alphad_\jb\hspace{-0.7pt})}+(c_1-1)\text{P}_{\!(\alpha(\alphad}
\CO_{\alpha_1\hspace{-0.8pt}\ldots\alpha_j);
\,\alphad_1\hspace{-0.8pt}\ldots\alphad_\jb\hspace{-0.7pt})},}[]
with similar expressions for the other primary operators.

With the results \eqref{QQbcs} and the zero component of \SCTwoPF we can
now compute the two-point functions
\eqna{\Vev{\Big[Q_{(\alpha}\Qb_{(\alphad}\CO_{\alpha_1\hspace{-0.8pt}
\ldots\alpha_j);\,
\alphad_1\hspace{-0.8pt}\ldots\alphad_\jb\hspace{-0.7pt})}\Big]_{p}(x)
\Big[\Qb_{\smash{(\betad}}Q_{\smash{(\beta}}
\COb_{\smash{\beta_1\hspace{-0.8pt}\ldots\beta_\jb\hspace{-0.7pt});\,
\betad_1\hspace{-0.8pt}\ldots\betad_j)}}\Big]_{p}(0)}&=\\
&\quad\hspace{-9cm}4C_\CO\frac{(j+2q) (\jb+2\qb) (j+\jb+2(q+\qb+1))}
{(j+1)^2(\jb+1)^2(j+\jb+2(q+\qb))} \frac{\xup_{(\alpha\betad}
\xup_{\alpha_1\betad_1}\!\!\cdots\xup_{\alpha_j)\betad_j}
\xup_{\vphantom{\betad}(\beta\alphad}\xup_{\vphantom{\betad}
\beta_1\alphad_1}\!\!\cdots\xup_{\vphantom{\betad}\beta_\jb
\hspace{-0.7pt})\alphad_{\jb}}}{x^{2(q+\qb+2)+j+\jb}},}[QQbSym]
\eqna{\Vev{\Big[Q^\alpha\Qb_{(\alphad}\CO_{\alpha\alpha_1\hspace{-0.8pt}
\ldots\alpha_{j-1};\,\alphad_1\hspace{-0.8pt}\ldots\alphad_\jb
\hspace{-0.7pt})}\Big]_{p}(x)\Big[\Qb^\betad Q_{\smash{(\beta}}
\COb_{\smash{\beta_1\hspace{-0.8pt}\ldots\beta_\jb\hspace{-0.7pt});\,
\betad\betad_1\hspace{-0.8pt}\ldots\betad_{j-1}}}\Big]_{p}(0)}&=\\
&\quad\hspace{-10cm}4C_\CO\frac{(j+1)(j-2(q-1))(\jb+2\qb)(j-\jb-2(q+\qb))}
{j(\jb+1)^2(j-\jb-2(q+\qb-1))}\frac{\xup_{(\alpha_1\betad_1}\!\!\cdots
\xup_{\alpha_{j-1})\betad_{j-1}}\xup_{\vphantom{\betad}
(\beta\alphad}\xup_{\vphantom{\betad}\beta_1\alphad_1}
\!\!\cdots\xup_{\vphantom{\betad}\beta_\jb\hspace{-0.7pt})\alphad_{\jb}}}
{x^{2(q+\qb+1)+j+\jb}},}[QQbAntiSymj]
\eqna{\Vev{\Big[Q_{(\alpha}\Qb^\alphad\CO_{\alpha_1\hspace{-0.8pt}\ldots
\alpha_j);\,\alphad\alphad_1\hspace{-0.8pt}\ldots\alphad_{\jb-1}}
\Big]_{p}(x)\Big[\Qb_{(\betad}Q^\beta\COb_{\smash{\beta\beta_1
\hspace{-0.8pt}\ldots\beta_{\jb-1};\,
\betad_1\hspace{-0.8pt}\ldots\betad_j)}}\Big]_{p}(0)}&=\\
&\quad\hspace{-10.2cm}4C_\CO\frac{(\jb+1)(\jb-2(\qb-1))(j+2q)
(j-\jb+2(q+\qb))}{\jb(j+1)^2(j-\jb+2(q+\qb-1))}\frac{
\xup_{\vphantom{\betad}(\alpha\betad}\xup_{\alpha_1\betad_1}\!\!\cdots
\xup_{\alpha_{j})\betad_{j}}\xup_{\vphantom{\betad}(\beta_1\alphad_1}
\!\!\cdots\xup_{\vphantom{\betad}\beta_{\jb-1})\alphad_{\jb-1}}}
{x^{2(q+\qb+1)+j+\jb}},}[QQbAntiSymjb]
\eqna{\Vev{\Big[Q^\alpha\Qb^\alphad\CO_{\alpha\alpha_1\hspace{-0.8pt}\ldots
\alpha_{j-1};\,\alphad\alphad_1\hspace{-0.8pt}\ldots\alphad_{\jb-1}}
\Big]_{p}(x)\Big[\Qb^\betad Q^\beta\COb_{\smash{\beta\beta_1\hspace{-0.8pt}
\ldots\beta_{\jb-1};\,\betad\betad_1\hspace{-0.8pt}\ldots\betad_{j-1}}}
\Big]_{p}(0)}&=\\
&\quad\hspace{-9cm}4C_\CO\frac{(j+1)(\jb+1)(j-2(q-1))(\jb-2(\qb-1)) (j+\jb
-2(q+\qb-1))}{j\jb(j+\jb-2(q+\qb-2))}\\
&\hspace{-3cm}\times\frac{\xup_{(\alpha_1\betad_1}
\!\!\cdots\xup_{\alpha_{j-1})\betad_{j-1}}\xup_{\vphantom{\betad}(\beta_1
\alphad_1}\!\!\cdots\xup_{\vphantom{\betad}\beta_{\jb-1})\alphad_{\jb-1}}}
{x^{2(q+\qb)+j+\jb}}.}[QQbAntiSymjjb]
As expected, \QQbAntiSymj and \QQbAntiSymjb are exchanged under
$(j,\jb\hspace{-0.7pt},q,\qb)\to(\jb\hspace{-0.7pt},j,\qb,q)$, while \QQbSym
and \QQbAntiSymjjb are invariant under
$(j,\jb\hspace{-0.7pt},q,\qb)\to(\jb\hspace{-0.7pt},j,\qb,q)$. When $j=\jb=0$,
only \QQbSym survives, while \QQbAntiSymj, \QQbAntiSymjb and \QQbAntiSymjjb
are defined when $j\ne0$, $\jb\ne0$ and $j\jb\ne0$ respectively.

\subsec{Orders \texorpdfstring{$\theta_1^2\thetab_1
\theta_2\thetab_2^{\hspace{0.7pt}2}$}{theta1sq thetabar1 theta2 thetabar2sq}
and \texorpdfstring{$\theta_1\thetab_1^{\hspace{0.7pt}2}\theta_2^2\thetab_2$}
{theta1 thetabar1sq theta2sq thetabar2}}
At this order we consider
\eqna{e^{i\theta Q+i\thetab
\Qb}\COind{\alpha}\big|_{\theta^2\thetab}&=-\frac{i}{4}\theta^2
\thetab^\alphad\Big\{\Big[Q^2\Qb_{(\alphad}
\CO_{\alpha_1\hspace{-0.8pt}\ldots\alpha_j;\,\alphad_1
\hspace{-0.8pt}\ldots\alphad_\jb\hspace{-0.7pt})}\Big]_p
+2c_5\text{P}^\alpha{}_{\!(\alphad}Q_{(\alpha}\CO_{\alpha_1
\hspace{-0.8pt}\ldots\alpha_j)
;\,\alphad_1\hspace{-0.8pt}\ldots\alphad_\jb\hspace{-0.7pt})}\\
&\hspace{5cm}-2c_6\frac{j}{j+1}\text{P}_{(\alpha_1(\alphad_1} Q^\alpha
\CO_{|\alpha\alpha_2\hspace{-0.5pt}\ldots\alpha_j);\,|\alphad\alphad_2
\hspace{-0.5pt}\ldots\alphad_\jb\hspace{-0.7pt})}
\Big\}\\
&\hspace{-0.5cm}+\frac{i}{4}\frac{\jb}{\jb+1}\theta^2\thetab_{(\alphad_1}
\Big\{\Big[Q^2\Qb^{\alphad}\CO_{\alpha_1\hspace{-0.8pt}\ldots\alpha_j;
\,|\alphad\alphad_2\ldots\alphad_\jb\hspace{-0.7pt})}\Big]_p
+2c_7\text{P}^{\alphad\alpha}Q_{(\alpha}\CO_{\alpha_1\hspace{-0.8pt}
\ldots\alpha_j);\,|\alphad\alphad_2
\hspace{-0.5pt}\ldots\alphad_\jb\hspace{-0.7pt})}\\
&\hspace{5.3cm}-2c_8\frac{j}{j+1}\text{P}^{\alphad}{}_{\!(\alpha_1}
Q^\alpha\CO_{|\alpha\alpha_2\hspace{-0.5pt}\ldots\alpha_j);\,|\alphad
\alphad_2\hspace{-0.5pt}\ldots\alphad_\jb\hspace{-0.7pt})}\Big\}.}[]
The $\theta_1^2\thetab_1\thetab_2$ order of \SCTwoPF arises because of the
descendants above via two-point functions of the form
$\vev{PQ\CO(x)\Qb\COb(0)}$. Using this we can determine
\begin{equation}
  \begin{aligned}
    c_5&=\frac{j+\jb-2(q-\qb-1)}{(\jb+1)(j-\jb-2(q+\qb-1))},&\qquad
    c_6&=\frac{j-\jb+2(q-\qb)}{(\jb+1)(j+\jb+2(q+\qb))},\\
    c_7&=\frac{j-\jb-2(q-\qb)}{\jb(j+\jb-2(q+\qb-2))}, &\qquad
    c_8&=\frac{j+\jb+2(q-\qb+1)}{\jb(j-\jb+2(q+\qb-1))}.
\end{aligned}
  \label{QsqQbcs}
\end{equation}
The results \eqref{QsqQbcs}, as well as \symQO and \antisymQO, allow us to
determine the two-point functions
\eqna{\Vev{\Big[Q^2\Qb_{(\alphad}\CO_{\alpha_1\hspace{-0.8pt}
\ldots\alpha_j;\,\alphad_1\hspace{-0.8pt}\ldots\alphad_\jb
\hspace{-0.7pt})}\Big]_p(x)\Big[\Qb^2 Q_{\smash{(\beta}}
\COb_{\smash{\beta_1\hspace{-0.8pt}\ldots\beta_\jb\hspace{-0.7pt});\,
\betad_1\hspace{-0.8pt}\ldots\betad_j}}\Big]_p(0)}=\\
&\hspace{-9cm}2^5iC_\CO(-1)^{j+\jb}\frac{(j+2q)(j-2(q-1))(\jb+2\qb)
(j+\jb+2(q+\qb+1))(j-\jb-2(q+\qb))}{(\jb+1)^2
(j+\jb+2(q+\qb))(j-\jb-2(q+\qb-1))}\\
&\times\frac{\xup_{(\alpha_1\betad_1}\!\!\cdots\xup_{\alpha_j)\betad_j}
\xup_{(\beta\alphad\vphantom{\betad}}
\xup_{\vphantom{\betad}\beta_1\alphad_1}\!\!\cdots
\xup_{\vphantom{\betad}\beta_\jb\hspace{-0.7pt})\alphad_{\jb}}}
{x^{2(q+\qb+2)+j+\jb}}}[QsqQbSym]
and
\eqna{\Vev{\Big[Q^2\Qb^\alphad\CO_{\alpha_1\hspace{-0.8pt}\ldots\alpha_j;\,
\alphad\alphad_1\hspace{-0.8pt}\ldots\alphad_{\jb-1}}\Big]_p(x)
\Big[\Qb^2Q^\beta\COb_{\smash{\beta\beta_1\hspace{-0.8pt}\ldots
\beta_{\jb-1};\,\betad_1\hspace{-0.8pt}\ldots\betad_j}}\Big]_p(0)}=\\
&\hspace{-10.1cm}2^5iC_\CO(-1)^{j+\jb}\frac{(\jb+1)(j+2q)(j-2(q-1))
(\jb-2(\qb-1))(j+\jb-2(q+\qb-1))(j-\jb+2(q+\qb))}{\jb
(j+\jb-2(q+\qb-2)) (j-\jb+2(q+\qb-1))}\\
&\hspace{-0.1cm}\times\frac{\xup_{(\alpha_1\betad_1}
\!\!\cdots \xup_{\alpha_j)\betad_j}\xup_{\vphantom{\betad}(\beta_1
\alphad_1}\!\!\cdots\xup_{\vphantom{\betad}\beta_{\jb-1})\alphad_{\jb-1}}}
{x^{2(q+\qb+1)+j+\jb}}.}[QsqQbAntiSym]

We can also obtain the two-point functions
\eqn{\Vev{\Big[\Qb^2 Q_{(\alpha}\CO_{\alpha_1\hspace{-0.8pt}
\ldots\alpha_j);\,\alphad_1\hspace{-0.8pt}\ldots\alphad_\jb}\Big]_p(x)
\Big[Q^2 \Qb_{\smash{(\betad}}\COb_{\smash{\beta_1\hspace{-0.8pt}\ldots
\beta_\jb;\,\betad_1\hspace{-0.8pt}\ldots\betad_j)}}
\Big]_p(0)}}[]
and
\eqn{\Vev{\Big[\Qb^2 Q^\alpha\CO_{\alpha\alpha_1\hspace{-0.8pt}
\ldots\alpha_{j-1};\,\alphad_1\hspace{-0.8pt}
\ldots\alphad_\jb}\Big]_p(x)\Big[Q^2\Qb^\betad
\COb_{\smash{\beta_1\hspace{-0.8pt}\ldots\beta_\jb;\,
\betad\betad_1\hspace{-0.8pt}\ldots\betad_{j-1}}}\Big]_p(0)}}[]
by letting $(j,\jb\hspace{-0.7pt},q,\qb)\to(\jb\hspace{-0.7pt},j,\qb,q)$ in
\QsqQbSym and \QsqQbAntiSym respectively.

\subsec{Order \texorpdfstring{$\theta_1^2\thetab_1^{\hspace{0.7pt}2}
\theta_2^2\thetab_2^{\hspace{0.7pt}2}$}{theta1sq thetabar1sq theta2sq
thetabar2sq}}
At this order we have to consider six new descendants:
\eqna{e^{i\theta Q+i\thetab\Qb}\COind{\alpha}\big|_{\theta^2\thetab^{2}}&=
\frac{1}{2^4}\theta^2\thetab^{\hspace{0.7pt}2}\Big\{\Big[Q^2\Qb^2
\COind{\alpha}\Big]_p
-4c_9\text{P}^{\alphad\alpha}\Big[Q_{(\alpha}\Qb_{(\alphad}
\CO_{\alpha_1\hspace{-0.8pt}\ldots\alpha_j);\,\alphad_1\hspace{-0.8pt}
\ldots\alphad_{\jb})}\Big]_p\\
&\hspace{-2.5cm}+4\frac{j}{j+1}c_{10}\text{P}^{\alphad}{}_{\!(\alpha_1}
\Big[Q^\alpha\Qb_{(\alphad}\CO_{|\alpha\alpha_2\hspace{-0.5pt}\ldots
\alpha_j);\,\alphad_1\hspace{-0.8pt}\ldots\alphad_\jb
\hspace{-0.7pt})}\Big]_p
-4\frac{\jb}{\jb+1}c_{11}\text{P}^\alpha{}_{\!(\alphad_1}\Big[
Q_{(\alpha}\Qb^{\alphad}\CO_{\alpha_1\hspace{-0.8pt}\ldots\alpha_j;\,|
\alphad\alphad_2\hspace{-0.5pt}\ldots\alphad_\jb\hspace{-0.7pt})}\Big]_p\\
&\hspace{-2.5cm}-4\frac{j\jb}{(j+1)(\jb+1)}c_{12}
\text{P}_{(\alpha_1(\alphad_1}\Big[Q^\alpha\Qb^\alphad
\CO_{|\alpha\alpha_2\hspace{-0.5pt}\ldots\alpha_j);\,|\alphad\alphad_2
\hspace{-0.5pt}\ldots\alphad_\jb\hspace{-0.7pt})}\Big]_p
-2^3c_{13}P^2\COind{\alpha}\\
&\hspace{-2.5cm}-2^5c_{14}\text{P}_{(\alpha_1(\alphad_1}
\text{P}^{\alphad\alpha}\CO_{|\alpha\alpha_2\hspace{-0.5pt}\ldots\alpha_j);
\,|\alphad\alphad_2\hspace{-0.5pt}\ldots\alphad_\jb
\hspace{-0.7pt})}\Big\}.}[]
The first four descendants result in terms $\theta_1^2
\thetab_1^{\hspace{0.7pt}2} \theta_2\thetab_2$ in the expansion of
\SCTwoPF, via two-point functions of the form $\vev{P[Q\Qb\CO]_p(x) [\Qb
Q\COb]_p(0)}$.  At this $\theta$-order these are not the only
contributions; two-point functions of the form
$\vev{P^2\CO(x)P^\dagger\COb(0)}$ also need to be taken into account. For
these contributions we need to first determine $c_{13,14}$. This can be
easily done since the associated descendants generate the order
$\theta_1^2\thetab_1^{\hspace{0.7pt}2}$ in the expansion of \SCTwoPF via
$\vev{P^2\CO(x)\COb(0)}$. After $c_{13,14}$ are determined, it is
straightforward to find $c_{9,\ldots,12}$. Note that in order to compute
$c_{9,\ldots,12}$ we also need the coefficients $c_{1,\ldots,4}$ in \eqref{QQbcs}.

Taking into consideration all the relevant contributions we find
\begin{equation}
  \begin{aligned}
    c_9&=\frac{j-\jb-2(q-\qb)}{j+\jb-2(q+\qb-2)},&\qquad
    c_{10}&=\frac{j+\jb+2(q-\qb+1)}{j-\jb+2(q+\qb-1)},\\
    c_{11}&=-\frac{j+\jb-2(q-\qb-1)}{j-\jb-2(q+\qb-1)},&\qquad
    c_{12}&=\frac{j-\jb+2(q-\qb)}{j+\jb+2(q+\qb)},\\
  \end{aligned}
  \label{QsqQbsqcsOne}
\end{equation}
and
\begin{equation}
  \begin{aligned}
    c_{13}&=\frac{\splitfrac{(j+\jb\hspace{-0.7pt})^2(j+\jb+2(q+\qb))
    -4(q-\qb)^2(j+\jb+2(q+\qb-2))+8(j+q)(\jb+\qb)}{+8(jq+\jb\qb+3q\qb)
    -4(j+\jb+2(q+\qb))}}{(j+\jb+2)(j+\jb+2(q+\qb))(j-\jb+2(q+\qb-1))
    (j-\jb-2(q+\qb-1))},\\
    c_{14}&=\frac{j(j+2)+\jb(\jb+2)-4(q(q-1)+\qb(\qb-1))}{(j+\jb+2(q+\qb))
    (j+\jb+2(q+\qb-1))(j+\jb-2(q+\qb-1))(j+\jb-2(q+\qb-2))}.
\end{aligned}
  \label{QsqQbsqcsTwo}
\end{equation}

Using \eqref{QsqQbsqcsOne} and \eqref{QsqQbsqcsTwo}, the zero component of
\SCTwoPF, as well as \QQbSym--\QQbAntiSymjjb, we can finally obtain
\eqna{\Vev{\Big[Q^2\Qb^2\COind{\alpha}\Big]_p(x)
\Big[\Qb^2Q^2\COindb{\beta}\Big]_p(0)}&=\\
&\hspace{-8.8cm}-2^8C_\CO\frac{\splitfrac{(j+2q)(j-2(q-1))(\jb+2\qb)
(\jb-2(\qb-1))(j-\jb+2(q+\qb))(j-\jb-2 (q+\qb))}
{\times(j+\jb+2 (q+\qb+1))(j+\jb-2 (q+\qb-1))}}
{(j+\jb+2 (q+\qb))(j-\jb+2 (q+\qb-1))
(j-\jb-2 (q+\qb-1))(j+\jb-2 (q+\qb-2))}\\
&\hspace{1.5cm}\times\frac{\xup_{(\alpha_1\betad_1}\!\!\cdots
\xup_{\alpha_j)\betad_j}\xup_{\vphantom{\betad}(\beta_1\alphad_1}
\!\!\cdots\xup_{\vphantom{\betad}\beta_\jb\hspace{-0.7pt})\alphad_{\jb}}}
{x^{2(q+\qb+2)+j+\jb}}.}[QsqQbsq]
This is the last two-point function to consider.

\newsec{Example: the supercurrent multiplet}[Example]
The supercurrent multiplet, or Ferrara--Zumino
multiplet~\cite{Ferrara:1974pz}, has $j=\jb=1$ and $q=\bar{q}=\frac{3}{2}$.
The shortening conditions $q=\frac{j}{2}+1$ and $\qb=\frac{\jb}{2}+1$ are
obviously satisfied, and, thus, as can be seen from Table
\ref{tab:Shortening-Unitary}, the multiplet can be expanded as
\begin{equation}
  \mathcal{J}_{\mu}(x,\theta,\bar{\theta})=j_{\mu}^{R}
  -\frac{i}{2}\bar{\sigma}_{\mu}^{\alphad\alpha}\theta^{\beta}Q_{(\beta}^
  {\phantom{R}}j_{\alpha)\alphad}^{R}+
  \frac{i}{2}\bar{\sigma}_{\mu}^{\alphad\alpha}\bar{\theta}^{\betad}
   \bar{Q}_{(\betad}j_{|\alpha|\alphad)}^{R}-
   \theta\sigma^{\nu}\bar{\theta}\,T_{\mu\nu}+\text{descendants},
\end{equation}
where
\eqn{T_{\mu\nu}=\tfrac14
\bar{\sigma}_{\mu}^{\alphad\alpha}\bar{\sigma}_\nu^{\betad\beta}
\left[Q_{(\beta\vphantom{\betad}}\Qb_{(\betad}\hspace{1pt}
j^R_{\alpha)\alphad)}\right]_p,}[]
which is obviously symmetric and traceless. In general, for any operator
with integer spin $\ell$, we have
$\CO_{\mu_{1}\hspace{-0.8pt}\ldots\mu_{\ell}}=
(-\frac{1}{2})^{\ell}\bar{\sigma}_{\mu_{1}}^{\alphad_{1}\alpha_{1}}
\!\cdots\bar{\sigma}_{\mu_{\ell}}^{\alphad_{\ell}\alpha_{\ell}}
\CO_{\alpha_{1}\hspace{-0.8pt}\ldots\alpha_{\ell};\,\dot{\alpha}_{1}
\hspace{-0.8pt}\ldots\dot{\alpha}_{\ell}}$.

The lowest component of $\mathcal{J}_{\mu}$ is the R-current, the
$\theta\bar{\theta}$ component is the energy-momentum tensor, while the
$\theta$, $\thetab$ components are the supersymmetry currents. A generic
supermultipet would contain many more primary component fields, but once
$(j,\jb\hspace{-0.7pt},q,\qb)\rightarrow(1,1,\frac{3}{2},\frac{3}{2})$ the
multiplet is shortened, and all other conformal primary components become
null. We check this by noting that all the component two-point functions
vanish in this limit except \eqref{SCTwoPF}, \eqref{symQO}, \eqref{symQOp}
and \eqref{QQbSym}, which become\foot{For two-point functions of operators
with four-vector indices see appendix~\ref{EtaForm},
section~\ref{intspin}.}
\begin{equation}
\langle j_{\mu}^{R}(x)j_{\nu}^{R}(0)\rangle
=\frac{c_{\mathcal{J}}}{2}\frac{1}{x^{6}}I_{\mu\nu}(x),
\end{equation}
\begin{equation}
  \langle
  Q^{\phantom{R}}_{(\alpha}j_{\alpha_{1}\hspace{-0.7pt})
  \dot{\alpha}_{1}}^{R}(x)\bar{Q}^{\phantom{R}}_{(\dot{\beta}}
  j_{|\beta_{1}|\dot{\beta}_{1}\hspace{-0.7pt})}^{R}(0)\rangle =
  ic_{\mathcal{J}}\frac{(\text{x}_{\alpha\dot{\beta}}
  \text{x}_{\alpha_{1}\dot{\beta}_{1}}+\text{x}_{\alpha_{1}\dot{\beta}}
  \text{x}_{\alpha\dot{\beta}_{1}})
  \text{x}_{\vphantom{\betad}\beta_{1}\dot{\alpha}_{1}}}{x^{10}},
\end{equation}
\begin{equation}
  \langle T_{\mu\nu}(x)T_{\rho\sigma}(0)\rangle =
  5c_{\mathcal{J}}\frac{1}{x^{8}}\left(I_{\mu\rho}(x)I_{\nu\sigma}(x)+
  I_{\mu\sigma}(x)I_{\nu\rho}(x)
  -\frac{1}{2}\eta_{\mu\nu}\eta_{\rho\sigma}\right),
\end{equation}
where
\eqn{I_{\mu\nu}(x)=\eta_{\mu\nu}-2\frac{x_{\mu}x_{\nu}}{x^{2}},}[Imunu]
and $c_{\mathcal{J}}>0$ in unitary theories.  These results imply
\eqn{\partial^{\mu}j_{\mu}^{R}=0,\qquad
\partial^{\mu}T_{\mu\nu}=0,\qquad
T_{\phantom{\mu}\!\mu}^{\mu}=0,\qquad
\partial_\mu(\bar{\sigma}^{\mu\betad\beta}
Q_{(\alpha}^{\phantom{R}}j_{\beta)\betad}^{R})=0,}[ConsCond]
which are the correct conservation conditions of $\mathcal{N}=1$
superconformal symmetry. Note that we did not need to impose these
conservation conditions as extra constraints. They follow from the correct
choice of quantum numbers of the multiplet.

Conservation conditions like \ConsCond hold for any conformal primary with
scaling dimension $\Delta=\frac{j+\jb}{2}+2$. As we just saw, the structure
of the superconformal two-point function forces them to hold, but note that
for three- and higher-point functions they are non-trivial Ward identities,
in the sense that they are not implied by the superconformal symmetry.

\newsec{Summary}[Summary]
Superconformal symmetry imposes powerful constraints on quantum field
theories in any dimensions. For four-dimensional $\mathcal{N}=1$ SCFTs the
form of superspace correlation functions consistent with the symmetry can
be obtained with various methods. In particular, superspace two-point
functions between superfields are determined up to an overall constant, and
the form of the three-point function is determined up to a few constants.
These results imply relations between the correlation functions involving
different components of the supermultiplet. These relations are physically
important, but have not been worked out explicitly in full generality.
For two-point functions of superconformal multiplets with quantum numbers
$(j,\jb,q,\qb)=(\ell,\ell,\Delta/2,\Delta/2)$ they were worked out
in~\cite{Poland:2010wg}, and the general results of this paper agree with
those obtained there.

In this work we developed a method that systematically computes such
relations based on superspace correlation functions. In particular we
decomposed the superspace two-point function to contributions from the
various conformal primaries and their descendants. Consequently, we
determined the relation imposed by the superconformal symmetry among the
two-point function coefficients of the conformal primary components in a
superconformal multiplet. This result enables us to determine all possible
shortening conditions associated with supermultiplets built from any
superconformmal primary operator. It also gives an alternative derivation
of the unitarity bounds. Our results are consistent with existing
literature.

The method described in this paper can also be applied to three-point
functions, which, together with the results presented here, will
systematically determine relations between OPE coefficients of conformal
primaries in a supermultiplet. This analysis will yield expressions for
superconformal blocks as linear combinations of conformal blocks.
Additionally, our formalism can be generalized to theories with more
supersymmetry and in other spacetime dimensions.

To make our calculation possible, we developed a \emph{Mathematica} package
that automates the simplification, expansion, and differentiation of
expressions built with four-vectors and two-component spinors. This package
can be useful in other calculations in supersymmetric field theories and
beyond.

\acknowledgments{We thank Jeff Fortin, Ken Intriligator, Zuhair Khandker,
Hugh Osborn, David Poland, and David Simmons-Duffin for helpful discussions
and comments. The research of DL is supported in part by the DOE grant
DE-FG-02-92ER40704. AS thanks the KITP for its hospitality during the
completion of this work. The research of AS is supported in part by the
National Science Foundation under Grant No.~1350180.}

\begin{appendices}

\newsec{The index-free formalism}[EtaForm]
The correlators of operators in generic Lorentz representations involve
many complicated tensor structures. The index-free formalism is an
efficient representation of these tensor structures. In addition, this
formalism also exposes the various linear relations between the tensor
structures. We thus employ this formalism in the implementation of our
calculation.  In this section, we introduce the index-free formalism that
describes superfields and correlators in $\mathcal{N}=1$ SCFTs.

The index-free formalism represents a symmetric tensor as a scalar
field defined on a space of auxiliary spinors,
\begin{equation}
  T_{(\alpha_{1}\hspace{-0.8pt}\ldots\alpha_{n})}\to\frac{1}{n!}
  \eta^{\alpha_{1}}\!\cdots\eta^{\alpha_{n}}T_{\alpha_{1}
  \hspace{-0.8pt}\ldots\alpha_{n}}
  \equiv T(\eta,n),
\end{equation}
where
\begin{equation}
  T_{(\alpha_{1}\hspace{-0.8pt}\ldots\alpha_{n})}\equiv
  \frac{1}{n!}\sum_{\mathcal{P}_{n}}\mathcal{P}_{n} T_{\alpha_{1}
  \hspace{-0.8pt}\ldots\alpha_{n}}
\end{equation}
is a symmetric tensor with $\mathcal{P}_{n}$ denoting permutations over the
indices $\alpha_{1},\ldots,\alpha_{n}$. Contrary to the spinor coordinates
of superspace, the auxiliary spinors $\eta$ commute with each other,
implying $\eta^{2}=\eta^{\alpha}\eta_{\alpha}
=\epsilon^{\alpha\beta}\eta_{\beta} \eta_{\alpha}=0$.  The index-free field
$T(\eta,n)$ constructed above can be mapped back to the traditional form by
differentiating with respect to the auxiliary spinors,
\begin{equation}
  \partial_{\eta^{\alpha_{1}}}\!\cdots\partial_{\eta^{\alpha_{n}}}T(\eta,n)
  =T_{(\alpha_{1}\hspace{-0.8pt}\ldots\alpha_{n})},
\end{equation}
where the spinor derivatives are defined in the usual way, i.e.\
$\partial_{\eta^{\alpha}}\eta^{\beta}\equiv\delta_{\alpha}{}^{\!\beta},
\partial_{\eta^{\alpha}}=-\epsilon_{\alpha\beta}\partial_{\eta_{\beta}}$.
Note the index-free form $T(\eta,n)$ contains exactly the same information
as the original traditional form $T_{(\alpha_{1}\hspace{-0.8pt}
\ldots\alpha_{n})}$. From now on we will omit the parentheses around
totally symmetrized indices.

An operator $\COind{\alpha}$ in the irrep $(j/2,\jb/2)$ of the Lorentz
group is represented by the index-free form
\eqn{\CO^\eta_{j,\hspace{1pt}\jb}\equiv\CO(\eta,j;\bar{\eta},\jb
\hspace{-0.7pt})\equiv
\frac{1}{j!\hspace{1pt}\jb!}\eta^{\alpha_{1}}\!
\cdots\eta^{\alpha_{j}}\bar{\eta}^{\alphad_{1}}\!\cdots
\bar{\eta}^{\alphad_{\jb}}\COind{\alpha}.}[COjjb]
The two-point function of any such conformal primary operator with its
conjugate is given by
\begin{equation}
  \langle\CO(\eta_{1},j;\bar{\eta}_{1},\jb\hspace{-0.7pt})(x)\,
  \CO(\eta_{2},j;\bar{\eta}_{2},\jb\hspace{-0.7pt})^{\dagger}(0)
  \rangle=C_{\CO}\frac{(\eta_{1}\text{x}\bar{\eta}_{2})^{j}
  (\eta_{2}\text{x}\bar{\eta}_{1})^{\jb}}{x^{2(q+\qb)+j+\jb}}.
  \label{eq:C2PF}
\end{equation}

We now turn to the supersymmetric case. Applying a generator $Q_{\alpha}$
on $\CO^\eta_{j,\hspace{1pt}\jb}$ will generate a reducible representation
of the Lorentz group, which contains two irreducible representations,
namely
\begin{equation}
  (Q\CO)^\eta_{j+1,\hspace{1pt}\jb}\equiv\frac{1}{j+1}[\eta Q,\CO^\eta_{j,
  \hspace{1pt}\jb}\},\qquad (Q\CO)^\eta_{j-1,\hspace{1pt}\jb}
  \equiv\frac{1}{j}[Q\partial_{\eta},\CO^\eta_{j,\hspace{1pt}\jb}\},
\end{equation}
where $Q$ acts with a commutator (resp.\ anticommutator) if $j+\jb$ is even
(resp.\ odd). With these normalizations we have
\begin{equation}
  \theta Q\CO_{j,\hspace{1pt}\jb}^\eta\equiv[\theta Q,
  \CO_{j,\hspace{1pt}\jb}^\eta\}
  =\theta\partial_{\eta}(Q\CO)^\eta_{j+1,\hspace{1pt}\jb}
  +\frac{j}{j+1}\theta\eta(Q\CO)^\eta_{j-1,\hspace{1pt}\jb}.
  \label{eq:QO}
\end{equation}
We can similarly write down each component of the superfield as a linear
combination of operators in irreps of the Lorentz group. We will suppress
the (anti)commutator and simply write the analogs of $\theta
Q\CO^\eta_{j,\hspace{1pt}\jb}$ from now on. For example, we have
\eqna{\mathcal{O}^\eta_{j,\hspace{1pt}\jb}\big|_{\theta\bar{\theta}}&=
-\theta\partial_{\eta}\,\bar{\theta}\partial_{\bar{\eta}}\,
((Q\bar{Q}\CO)^\eta_{j+1,\hspace{1pt}\jb+1;\hspace{1pt}p}
-ic_{1}\eta\partial_{x}\bar{\eta}\,\CO^\eta_{j,\hspace{1pt}\jb})\\
&\quad-\frac{j}{j+1}\theta\eta\,\bar{\theta}\partial_{\bar{\eta}}\,
((Q\bar{Q}\CO)^\eta_{j-1,\hspace{1pt}\jb+1;\hspace{1pt}p}-
ic_{2}\,\partial_{\eta}\partial_{x}\bar{\eta}\,
\CO^\eta_{j,\hspace{1pt}\jb})\\
&\quad+\frac{\jb}{\jb+1}\theta\partial_{\eta}\,\bar{\theta}\bar{\eta}\,
((Q\bar{Q}\CO)^\eta_{j+1,\hspace{1pt}\jb-1;\hspace{1pt}p}+
ic_{3}\,\eta\partial_{x}\partial_{\bar{\eta}}\,
\CO^\eta_{j,\hspace{1pt}\jb})\\
&\quad+\frac{j\jb}{(j+1)(\jb+1)}\theta\eta\,\bar{\theta}\bar{\eta}\,
((Q\bar{Q}\CO)^\eta_{j-1,\hspace{1pt}\jb-1;\hspace{1pt}p}-ic_{4}\,
\partial_{\eta}\partial_{x}\partial_{\bar{\eta}}\,\CO^\eta_{j,
\hspace{1pt}\jb}),}[]
\eqna{\mathcal{O}^\eta_{j,\hspace{1pt}\jb}\big|_{\theta^{2}\bar{\theta}}
&=-\frac{i}{4}\theta^{2}\,\bar{\theta}\partial_{\bar{\eta}}\left((Q^{2}
\bar{Q}\CO)^\eta_{j,\hspace{1pt}\jb+1;\hspace{1pt}p}+2ic_{5}\,
\partial_{\eta}\partial_{x}\bar{\eta}\,(Q\CO)^\eta_{j+1,\hspace{1pt}\jb
\hspace{1pt};\hspace{1pt}p}-2ic_{6}\frac{j}{j+1}\,\eta\partial_{x}
\bar{\eta}\,(Q\CO)^\eta_{j-1,\hspace{1pt}\jb;\hspace{1pt}p}
\right)\\
&\quad\hspace{-0.5cm}+\frac{i}{4}\frac{\jb}{\jb+1}\theta^{2}\,
\bar{\theta}\bar{\eta}\left((Q^{2}\bar{Q}\CO)^\eta_{j,\hspace{1pt}
\jb-1;\hspace{1pt}p}+2ic_{7}\,\partial_{\eta}\partial_{x}
\partial_{\bar{\eta}}\,(Q\CO)^\eta_{j+1,\hspace{1pt}\jb;
\hspace{1pt}p}-2ic_{8}\frac{j}{j+1}\,\eta\partial_{x}\partial_{\bar{\eta}}
\,(Q\CO)^\eta_{j-1,\hspace{1pt}\jb;\hspace{1pt}p}\right),}[]
and
\eqna{\mathcal{O}^\eta_{j,\hspace{1pt}\jb}\big|_{\theta^{2}\bar{\theta}^
{2}}
&=\frac{1}{2^4}\theta^{2}\bar{\theta}^{\hspace{1pt}2}\!\left(\!
(Q^{2}\bar{Q}^{2}\CO)^\eta_{j,\hspace{1pt}\jb;\hspace{1pt}p}
\!-4ic_{9}\,\partial_{\eta}\partial_{x}\partial_{\bar{\eta}}\,
(Q\bar{Q}\CO)^\eta_{j+1,\hspace{1pt}\jb+1;\hspace{1pt}p}
\!+4\frac{j}{j+1}ic_{10}\,\eta\partial_{x}\partial_{\bar{\eta}}\,
(Q\bar{Q}\CO)^\eta_{j-1,\hspace{1pt}\jb+1;\hspace{1pt}p}\right. \\
&\quad-4\frac{\jb}{\jb+1}ic_{11}\,\partial_{\eta}\partial_{x}\bar{\eta}\,
(Q\bar{Q}\CO)^\eta_{j+1,\hspace{1pt}\jb-1;\hspace{1pt}p}
-4\frac{j\jb}{(j+1)(\jb+1)}ic_{12}\,\eta\partial_{x}\bar{\eta}\,
(Q\bar{Q}\CO)^\eta_{j-1,\hspace{1pt}\jb-1;\hspace{1pt}p}\\
&\quad\left.\vphantom{\frac{j}{j+1}}+2^3c_{13}\,
\partial_x^{2}\CO^\eta_{j,\hspace{1pt}\jb}+2^5c_{14}\,
\eta\partial_{x}\bar{\eta}\,\partial_{\eta}\partial_{x}
\partial_{\bar{\eta}}\,\CO^\eta_{j,\hspace{1pt}\jb}\right),}[]
in accord with corresponding expressions in section \TwoPFs.

The superfield two-point function of $\CO^\eta_{j,\hspace{1pt}\jb}$,
following \SCTwoPF, is given by
\begin{equation}
  \langle \mathcal{O}^{\eta_{1}}_{j,\hspace{1pt}\jb}(z_1)
  (\mathcal{O}^{\eta_2}_{j,\hspace{1pt}\jb})^{\dagger}(z_2)\rangle
  =C_{\mathcal{O}}\frac{(\eta_{1}\text{x}_{1\bar{2}}\bar{\eta}_{2})^{j}
  (\eta_{2}\text{x}_{\bar{1}2}\bar{\eta}_{1})^{\jb}}{\osbx{1\bar{2}}{2q+j}
  \osbx{\bar{1}2}{2\qb+\jb}}.
  \label{eq:SF2PF}
\end{equation}
By conformal symmetry, the two-point function coefficient of all conformal
primary operators in a supermultiplet is proportional to $C_\CO$, with
coefficient determined by the quantum numbers
$(j,\jb\hspace{-0.7pt},q,\bar{q})$ of the lowest component. The
coefficients $c_{i}$ of the descendants are also determined by this
information. In this paper we use the superfield two-point function
\eqref{eq:SF2PF} to explicitly work out all these coefficients.

For example, to determine the two-point function coefficients of
$(Q\CO)^\eta_{j+1,\hspace{1pt}\jb}$ and
$(Q\CO)^\eta_{j-1,\hspace{1pt}\jb}$, we simply expand (\ref{eq:SF2PF}) and
match to the expected form obtained with (\ref{eq:C2PF}) and (\ref{eq:QO}).
In particular, we can define
\begin{equation}
  \langle (Q\CO)^{\eta_1}_{j\pm1,\hspace{1pt}\jb}(x)((Q\CO)^{\eta_2}_
  {j\pm1,\hspace{1pt}\jb})^{\dagger}(0)\rangle =C_{(Q\mathcal{O})_{j\pm1,
  \hspace{1pt}\jb}}\frac{(\eta_{1}\text{x}\hspace{0.5pt}
  \bar{\eta}_{2})^{j\pm1}(\eta_{2}\text{x}
  \hspace{0.5pt}\bar{\eta}_{1})^{\jb}}{x^{2(q+\qb)+1+j+\jb\pm1}},
\end{equation}
and obtain (setting $x_2=0$)
\begin{align}
  \langle \mathcal{O}^{\eta_1}_{j,\hspace{1pt}\jb}
  (z_1)(\mathcal{O}^{\eta_2}_{j,\hspace{1pt}\jb})^{\dagger}(z_2)\rangle
  \big|_{\theta_{1}\bar{\theta}_{2}}&=
  C_{(Q\mathcal{O})_{j+1,\hspace{1pt}\jb}}\,\theta_{1}\partial_{\eta_{1}}\,
  \bar{\theta}_{2}\partial_{\bar{\eta}_{2}}\,\frac{(\eta_{1}\text{x}
  \hspace{0.5pt}\bar{\eta}_{2})^{j+1}(\eta_{2}\text{x}
  \hspace{0.5pt}\bar{\eta}_{1})^{\jb}}{x^{2(q+\qb+1)+j+\jb}}\\
  &\quad+\left(\frac{j}{j+1}\right)^{2}C_{(Q\mathcal{O})_{j-1,
  \hspace{1pt}\jb}}\,\theta_{1}\eta_{1}\,\bar{\theta}_{2}\bar{\eta}_2\,
  \frac{(\eta_{1}\text{x}\hspace{0.5pt}\bar{\eta}_{2})^{j-1}
  (\eta_{2}\text{x}\hspace{0.5pt}\bar{\eta}_{1})^{\jb}}{x^{2(q+\qb)+j+\jb}
  }.
\end{align}
From this relation we can determine the two unknowns
$C_{(Q\mathcal{O})_{j\pm1, \hspace{1pt}\jb}}$.  There will be two
independent tensor structures appearing on both sides, providing the two
necessary equations.  The result is
\begin{equation}
C_{(Q\mathcal{O})_{j+1,\hspace{1pt}\jb}}=2i\frac{j+2q}{(j+1)^{2}}
C_{\mathcal{O}},\qquad C_{(Q\mathcal{O})_{j-1,\hspace{1pt}\jb}}
=2i\frac{(j+1)(j-2(q-1))}{j}C_{\mathcal{O}}.
\end{equation}

We use similar methods to determine coefficients appearing in all
components of the general superconformal primary superfield
$\mathcal{O}^\eta_{j,\hspace{1pt}\jb}$. In some cases, in order to obtain
the primary two-point function coefficients, we need to determine the
descendant coefficients first. For example, $c_{1,\ldots,4}$ can be
determined through the $\theta_{1}\bar{\theta}_{1}$ or $\theta_2\thetab_2$
order of (\ref{eq:SF2PF}), which is then used as input for determining
$C_{(Q\bar{Q}\mathcal{O})_{j\pm1,\hspace{1pt}\jb\pm1}}$ in the
$\theta_{1}\bar{\theta}_{1}\theta_{2}\bar{\theta}_{2}$ order of
(\ref{eq:SF2PF}). In general, the number of independent tensors in a
particular order may exceed the number of unknown coefficients.  These
extra constrains provide non-trivial consistency checks for our results.

\subsec{Operators with integer spin}[intspin]
Operators $\COind{\alpha}$ with $j=\jb=\ell$ form an interger spin
representation of the Lorentz group. We can convert the spinor indices to
four-vector indices by
\begin{equation}
  \mathcal{O}_{\mu_{1}\hspace{-0.8pt}\ldots\mu_{\ell}}\equiv
  (-\tfrac{1}{2})^{\ell}\bar{\sigma}_{\mu_{1}}^{\alpha_{1}
  \dot{\alpha}_{1}}\!\cdots\bar{\sigma}_{\mu_{\ell}}^{\alpha_{\ell}
  \dot{\alpha}_{\ell}}\mathcal{O}_{\alpha_{1}\hspace{-0.8pt}\ldots
  \alpha_{\ell};\,\dot{\alpha}_{1}\hspace{-0.8pt}\ldots
  \dot{\alpha}_{\ell}}.\label{eq:ConvertionSpinorVectorIndex}
\end{equation}
Then, $\mathcal{O}_{\mu_{1}\hspace{-0.8pt}\ldots\mu_{\ell}}$ is a symmetric
traceless tensor. For completeness, we derive in this section the explicit
form of the two-point function of $\mathcal{O}_{\mu_{1}\hspace{-0.8pt}
\ldots\mu_{\ell}}$ from (\ref{eq:C2PF}). We first rewrite $\mathcal{O}$ in
an index-free form,
\begin{equation}
  \mathcal{O}_{\ell}^b=\frac{1}{\ell\hspace{0.5pt}!}b^{\mu_{1}}\!\cdots
  b^{\mu_{\ell}}\mathcal{O}_{\mu_{1}\hspace{-0.8pt}\ldots\mu_{\ell}},
  \label{eq:IndexFreeVector}
\end{equation}
where $b^{\mu_{i}}$ are auxiliary bosonic four-vectors satisfying
$b^{\mu}b_{\mu}=0$, corresponding to the fact that $\mathcal{O}$ is
traceless. Then,
\begin{equation}
  \mathcal{O}_{\mu_{1}\hspace{-0.8pt}\ldots\mu_{\ell}}=
  \frac{\partial}{\partial b^{\mu_1}}\cdots\frac{\partial}{\partial
  b^{\mu_\ell}}\mathcal{O}_{\ell}^b-\text{traces}.
  \label{eq:FreeIndexVector}
\end{equation}
Equation (\ref{eq:ConvertionSpinorVectorIndex}) then implies the following
mapping from \COjjb to (\ref{eq:IndexFreeVector}):
\begin{equation}
  \mathcal{O}_{\ell}^b=(-\tfrac12)^{\ell}\frac{1}{\ell
  \hspace{0.5pt}!}(\partial_{\eta}\text{b}\hspace{0.5pt}
  \partial_{\bar{\eta}})^{\ell}
  \mathcal{O}^\eta_{j=\ell,\hspace{1pt}\jb=\ell},\qquad
  \text{b}_{\alpha\alphad}=\sigma^\mu_{\alpha\alphad}b_\mu.
\end{equation}
We apply this mapping to the two-point function \eqref{eq:SF2PF} and get
\eqn{\vev{\mathcal{O}^{b_1}_{\ell}(x)\,
(\mathcal{O}_{\ell}^{b_2})^{\dagger}(0)}=
c_{\CO}\frac{\ell\hspace{0.5pt}!}{2^{\ell}}
\frac{(x^{2}b_{1}\cdot b_{2}-2(b_{1}\cdot x)(b_{2}\cdot
x))^{\ell}}{x^{2(\Delta_{\CO}+\ell)}},}[]
where we have used $b_{1,2}^{2}=0$. The coefficient
$c_\mathcal{O}=(-1)^\ell C_\mathcal{O}$ is positive in unitary theories.
Finally, using (\ref{eq:FreeIndexVector}), we get the familiar result
\begin{equation}
  \langle\mathcal{O}_{\mu_{1}\hspace{-0.8pt}
  \ldots\mu_{\ell}}^{\phantom{\dagger}}(x)\,
  \mathcal{O}_{\nu_{1}\hspace{-0.8pt}\dots\nu_{\ell}}^{\dagger}(0)
  \rangle=c_{\CO}\frac{(\ell\hspace{0.5pt}!)^{3}}{2^{\ell}}
  \frac{I_{(\mu_{1}|\nu_{1}\hspace{-0.5pt}|}(x)\cdots
  I_{\mu_{\ell})\nu_{\ell}}(x)-\text{traces}}{x^{2\Delta_{\CO}}}.
\end{equation}

\newsec{The \emph{Mathematica} package}[Package]
In this appendix we provide more details on our \emph{Mathematica}\foot{In
the development of the package we have used \emph{Mathematica} 9.} package,
which is an efficient tool for expanding functions with Grassmann variables
and simplifying expressions with Lorentz structures.

This package handles general expressions built with any number of
four-vectors $x^{\mu}$ and two-component spinors $\theta_{\alpha}$,
$\bar{\theta}_{\dot{\alpha}}$ and $\eta_{\vphantom{\betad}\beta}$,
$\bar{\eta}_{\dot{\beta}}$, where $\theta$ and $\bar{\theta}$ are Grassmann
variables, while $\eta$ and $\bar{\eta}$ are commuting variables. We use
the index-free formalism of the previous appendix to represent all
expressions with free indices as a Lorentz-invariant scalar function. The
standard simplifications for such expressions are automated, employing
rules like $\theta_{i\hspace{0.7pt}\alpha}\theta_{i \hspace{0.7pt}\beta}
\theta_{i \hspace{0.7pt}\gamma}=0$, $\theta_{\alpha}\theta^{\beta}=
-\frac{1}{2}\delta_{\alpha}^{\phantom{\alpha}\!\beta}\theta^{2}$,
$x_{\mu}x_{\nu}\,\eta_{1}\sigma^{\mu}\bar{\sigma}^{\nu}\eta_{2}=
-x^{2}\,\eta_{1}\eta_{2}$, etc. In the package we follow the conventions of
Wess \& Bagger \cite{Wess:1992cp}.

One of the main features of our package is the implementation of the Taylor
expansion in the Grassmann variables.  For example, given a function $f$
involving two undotted Grassmann spinors
$\theta_{1,2\hspace{0.7pt}\alpha}$, the program decomposes it as follows:
\eqn{
f(x,\theta_{1},\theta_{2})=f^{(0,0)}(x)+f_{\alpha}^{\left(1,0\right)}(x)
\theta_{1}^{\alpha}+f_{\beta}^{\left(0,1\right)}(x)\theta_{2}^{\beta}
+f_{\alpha\beta}^{(1,1)}(x)\theta_{1}^{\alpha}\theta_{2}^{\beta}
+f^{\left(2,0\right)}(x)\theta_{1}^{2}+f^{\left(0,2\right)}(x)
\theta_{2}^{2},}[]
where $x$ refers to any other bosonic variables $f$ depends on.  When more
copies of Grassmann variables $\theta_{i\hspace{0.7pt}\alpha}$ and
$\bar{\theta}_{i\hspace{0.7pt}\dot{\alpha}}$ appear, the size of the
computation quickly grows beyond human capability. In this work, we
decomposed a generic superconformal two-point function which depends on two
pairs of $\theta$, $\bar{\theta}$ and two pairs of $\eta$, $\bar{\eta}$.
The fully simplified result still contains $\sim 100$ distinct tensor
structures.  This process takes about 7 seconds on a laptop computer.

In addition, we implement generic differential operators such as
$\partial_{x}^{2}$,
$\partial_{\eta_{1}}\sigma_{\mu}\partial_{x}^\mu\hspace{0.7pt}
\bar{\eta}_{2}$, $\partial_{\eta_{1}}^{\alpha}
(\partial_{\eta_{2}})_\alpha$, etc. They are used to work out the
contributions from particular descendant operators to the superconformal
two-point function.  This involves acting with up to eight such operators
on a generic conformal two-point function, which takes about a minute to
complete.

We hope that this package will help realize complicated calculations both
in supersymmetric field theories and beyond.

\end{appendices}

\newpage
\bibliography{SC_2PFs}

\end{document}